\documentclass[onecolumn,pre,superscriptaddress, longbibliography, nofootinbib, notitlepage]{revtex4-1}

\usepackage{bbm}
\usepackage{url}

\usepackage{amsmath,amsfonts,amssymb}
\usepackage{graphicx}
\usepackage{subcaption}
\usepackage{color}

\newcommand{\E}{\mathbb{E}}
\newcommand{\mgs}{\mathrm{mgs}}
\newcommand{\gs}{\mathrm{gs}}
\newcommand{\ms}{\mathrm{ms}}
\newcommand{\rumgs}{r^u_{\mgs}}
\newcommand{\rlmgs}{r^l_{\mgs}}
\newcommand{\rugs}{r^u_{\gs}}
\newcommand{\rlgs}{r^l_{\gs}}
\newcommand{\rums}{r^{<\max}_{\ms}}
\newcommand{\rlms}{r^{>\min}_{\ms}}

\begin{document}

\title{Network Constraints on the Mixing Patterns of Binary Node Metadata}
\author{Matteo Cinelli}
\affiliation{Ca' Foscari University of Venice, Department of Environmental Sciences, Informatics and Statistics, 30172, Mestre (VE), Italy}
\affiliation{Applico Lab, CNR-ISC, 00185, Rome, Italy}

\author{Leto Peel}
\affiliation{Institute of Data Science, Faculty of Science and Engineering, Maastricht University}
\affiliation{Department of Data Analytics and Digitalisation,
School of Business and Economics, Maastricht University}

\author{Antonio Iovanella}
\affiliation{University of Rome ``Tor Vergata'', Via del Politecnico 1, Rome, Italy}

\author{Jean-Charles Delvenne}
\affiliation{ICTEAM, Universit\'{e} catholique de Louvain, Louvain-la-Neuve, Belgium}
\affiliation{CORE, Universit\'{e} catholique de Louvain, Louvain-la-Neuve, Belgium}

\begin{abstract}
We consider the network constraints on the bounds of the assortativity coefficient, which aims to quantify the tendency of nodes with the same attribute values to be connected. The assortativity coefficient can be considered as the Pearson's correlation coefficient of node metadata values across network edges and lies in the interval $[-1,1]$. However, properties of the network, such as degree distribution and the distribution of node metadata values place constraints upon the attainable values of the assortativity coefficient. This is important as a particular value of assortativity may say as much about the network topology as about how the metadata are distributed over the network --  a fact often overlooked in literature where the interpretation tends to focus simply on the propensity of similar nodes to link to each other, without any regard on the constraints posed by the topology. In this paper we quantify the effect that the topology has on the assortativity coefficient in the case of binary node metadata. Specifically we look at the effect that the degree distribution, or the full topology, and the proportion of each metadata value has on the extremal values of the assortativity coefficient. We provide the means for obtaining bounds on the extremal values of assortativity for different settings and demonstrate that under certain conditions the maximum and minimum values of assortativity are severely limited, which may present issues in interpretation when these bounds are not considered.
\end{abstract}

\maketitle

\section{Introduction}
\label{Intro}
Assortative mixing is the tendency of nodes with similar  attribute values (also referred to as \textit{node metadata}) to be connected to each other in a network. 
For example, in a human social network, where nodes are people and edges are interactions between them, node metadata might include the ages or genders of those people. 
Newman's assortativity coefficient, $r$,~\cite{newman2003mixing} was proposed as a means to quantify the level of assortative mixing in a network with respect to a particular piece of node metadata. This assortativity coefficient can be considered as the Pearson's correlation coefficient of node metadata values across edges. 
Just as correlation plays an important role in identifying relationships between pairs of variables, assortativity plays a fundamental role in understanding how a network is organized with respect to a given attribute of the nodes. 
However, there are issues that may arise when interpreting assortativity values as simply a measure of propensity of adjacent nodes to have similar metadata values. 
Here we consider one of these issues, specifically that the properties of the network (irrespective of how the metadata values are assigned to nodes) and the properties of the node metadata (irrespective of how the nodes are connected) can each influence the range of attainable assortativity values. A trivial illustrative example is the following: suppose we know that an undirected network is connected and the node metadata take value $1$ for some nodes and $0$ for the others. Then we know that assortativity $r=+1$ is unattainable because there will be at least one edge with metadata values $(0,1)$. If the graph is not bipartite, we know that assortativity $r=-1$ is unattainable because there will be at least one edge with metadata values $(0,0)$ or $(1,1)$.
As a consequence the assortativity coefficient conflates information of the network topology with how the metadata values are distributed over the network.

Pearson's correlation coefficient is used to measure the association between two variables $\mathbf{x}$ and $\mathbf{y}$ (which do not typically come from a network). 
Each pair $(x_t, y_t)$ is assumed to be sampled from the same joint probability distribution, independent of any other pair $(x_{t'}, y_{t'})$, where $t' \neq t$. We apply Pearson's correlation coefficient to calculate assortativity by treating the metadata ($c_i=x_t, c_j=y_t$) of two nodes, $i$ and $j$, connected by an edge $t=(i,j)$ as a sample $(x_t, y_t)$.
Since assortativity is an application of Pearson's correlation coefficient it inherits the same potential issue of a reduced interval.
Even when $\mathbf{x}$ and $\mathbf{y}$ are identically distributed, the underlying network structure can also limit the attainable values of assortativity, in that it limits the way pairs $(x_t,y_t)$ can be formed.

Each sample pair $(x_t, y_t)$ is assumed to be sampled from the same joint probability distribution, independent of any other pair $(x_{t'}, y_{t'})$, where $t' \neq t$. We apply Pearson's correlation coefficient to calculate assortativity by treating the metadata $c_i=x_t, c_j=y_t$ of two nodes, $i$ and $j$, connected by an edge $t=(i,j)$ as a sample $(x_t, y_t)$. A known, but often overlooked, issue is that the range of Pearson's correlation coefficient can be smaller than the usual reference interval $[- 1, 1]$. This reduced interval occurs, for instance, when the marginal distributions of $\mathbf{x}$ and $\mathbf{y}$ are not equal (aside from differences in scale and location)~\cite{frechet1960tableaux, hoeffding1994scale}.  

\begin{figure}
  \centering
   \includegraphics[trim=0cm 0.5cm 0cm 0.5cm, clip=true,width=.5\linewidth]{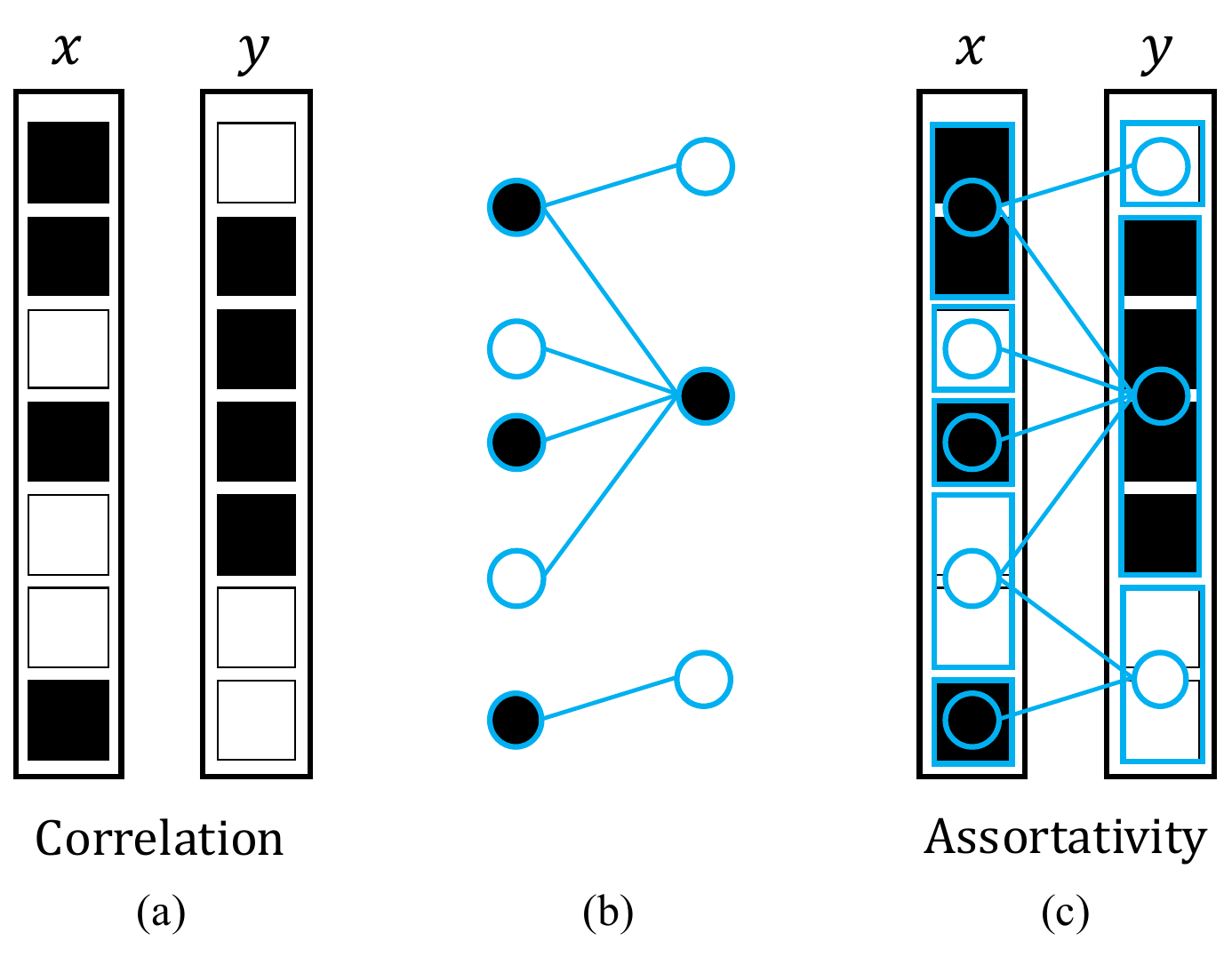}
\caption{Assortativity as correlation in a network. Panel (a): For the correlation of two binary variables, $x$ and $y$, we assume each row corresponds to a draw from a simple joint distribution $\textrm{P}(x, y)$. For example, each row is a person, $x$ indicates gender \{$male, female$\} and $y$ indicates if they wear glasses \{$yes, no$\}. Panel (b): For assortativity we just have one binary variable $c$ (e.g., gender) and pairs of variables connected by an edge in the network, which for illustrative purposes we can consider directed from left to right. Panel (c): Now each node (and therefore its gender) will appear in as many rows as it has edges, so we should no longer consider these to be samples from the same bivariate distribution $\textrm{P}(x, y)$ because it does not account for the network structure, i.e., a node cannot be female in one interaction and male in another interaction. To account for network structure, we instead consider each pair as a sample from a distribution over node pairs in the adjacency matrix. Changing the model in this way does not change the value of the assortativity coefficient, but it does change range of attainable values due to the added constraints of the network structure.}
   \label{correlation_assortativity}
 \end{figure}

Since assortativity is an application of Pearson's correlation coefficient it inherits the same potential issue of a reduced interval.
Even when $\mathbf{x}$ and $\mathbf{y}$ are identically distributed, the underlying network structure can also limit the attainable values of assortativity, as the network structure limits the way pairs $(x_t,y_t)$ can be formed.
Figure~\ref{correlation_assortativity} shows an example, illustrating that the presence of a (directed) network limits the possible metadata pairs we may observe. For instance, if node $i$ is incident upon $k_i$ edges, then node $i$'s metadata value $c_i$ appears in all $k_i$ pairs that represent its edges. This dependence on the degree sequence further limits the bounds of attainable assortativity, beyond simply considering the proportions of the metadata values. 
 
Here we focus on calculating the bounds of assortativity in the special case in which the graph is undirected (i.e., edge $(i,j)$ can equivalently be written as $(j,i)$) and unweighted, without any self loops or multiedges, and the node metadata are restricted to binary values, e.g., gender of actors in social networks. 
This setting is relevant, for example, when studying important phenomena such as gender homophily~\cite{jadidi2017gender} and related perception biases in social networks~\cite{lerman2016majority, lee2017homophily}.
We provide methods to calculate bounds on the attainable range of assortativity under the assumption that specific properties of the network are fixed. We focus on two types of properties: those relating to the network structure (specific degree sequence or specific graph topology) and those relating to the node metadata (proportion of nodes per category or specific assignment of nodes to categories). Considering all possible combinations of these network and metadata properties provides us with three different spaces of configurations (omitting the fourth combination as it corresponds to just the single network configuration that we observe):
\begin{enumerate}
\item the \textit{metadata-graph space} (mgs) -- the ensemble of configurations with a given degree sequence and proportion of metadata values
\item the \textit{graph space} (gs) -- the ensemble of configurations with a given degree sequence and specific node metadata assignment 
\item  the \textit{metadata space} (ms) -- the ensemble of configurations with a specific topology and proportion of metadata values. %
\end{enumerate}
In the metadata-graph space the range of assortativity can be explored by computing the assortativity coefficient over the set of all possible graphs with the observed degree sequence (the graph space) combined with set of all possible permutations of the metadata vector (the metadata space), i.e., a vector $\mathbf{c}$ in which each entry $c_i$ represents the metadata value of node $i \in \{1,...,n\}$ in the network.  For this space, we present combinatorial bounds on the largest possible range of attainable assortativity values since it contains both the graph space and the metadata space. Thus, bounding assortativity in the metadata-graph space means bounding assortativity with respect to all the possible values it can assume within the other spaces. 
Similarly for the graph space~\cite{fosdick2016configuring} we present combinatorial bounds for the range of assortativity, $r$, for all possible configurations of the observed degree sequence, but this time with the metadata vector fixed. 
Finally, the range of assortativity in the metadata space is harder to bound analitycally, being strictly dependent on the topology of the network. Instead we explore the range via a complete enumeration (when computationally feasible~\cite{park2007distribution}) of all possible permutations of the metadata vector. For larger graphs we resort to heuristic methods.

\begin{figure}[b]
\centering
\begin{subfigure}{0.6\linewidth}
   \centering
   \includegraphics[trim=0cm 1.5cm 0cm 0cm, clip=true, width=\linewidth]{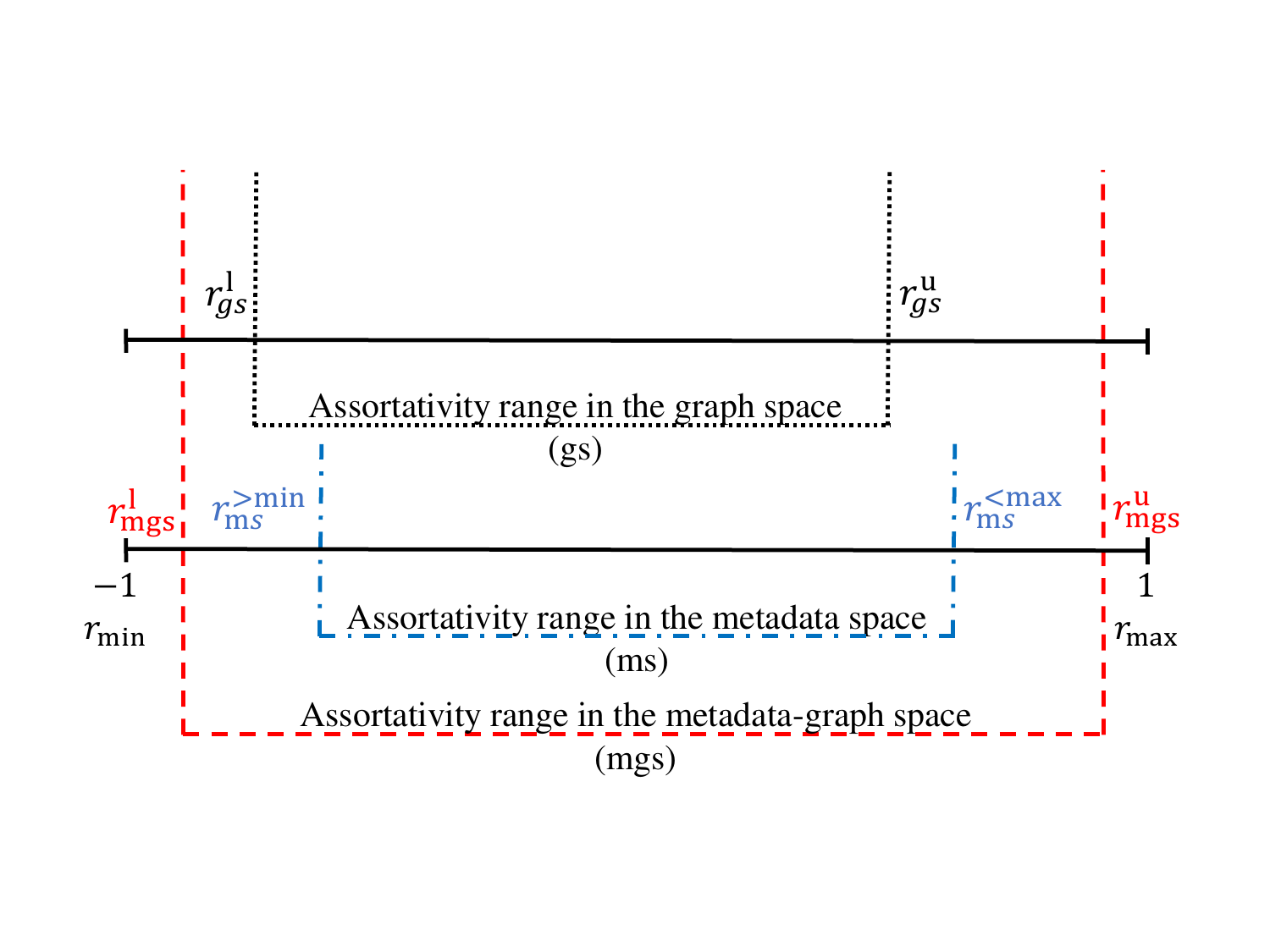}
   \caption{}
   \label{fig:dragratio2}
\end{subfigure}
\\[\baselineskip]
\begin{subfigure}[h]{0.45\linewidth}
   \centering
   \includegraphics[scale=.4, trim=2.6cm 8cm 0cm 1cm, clip=true]{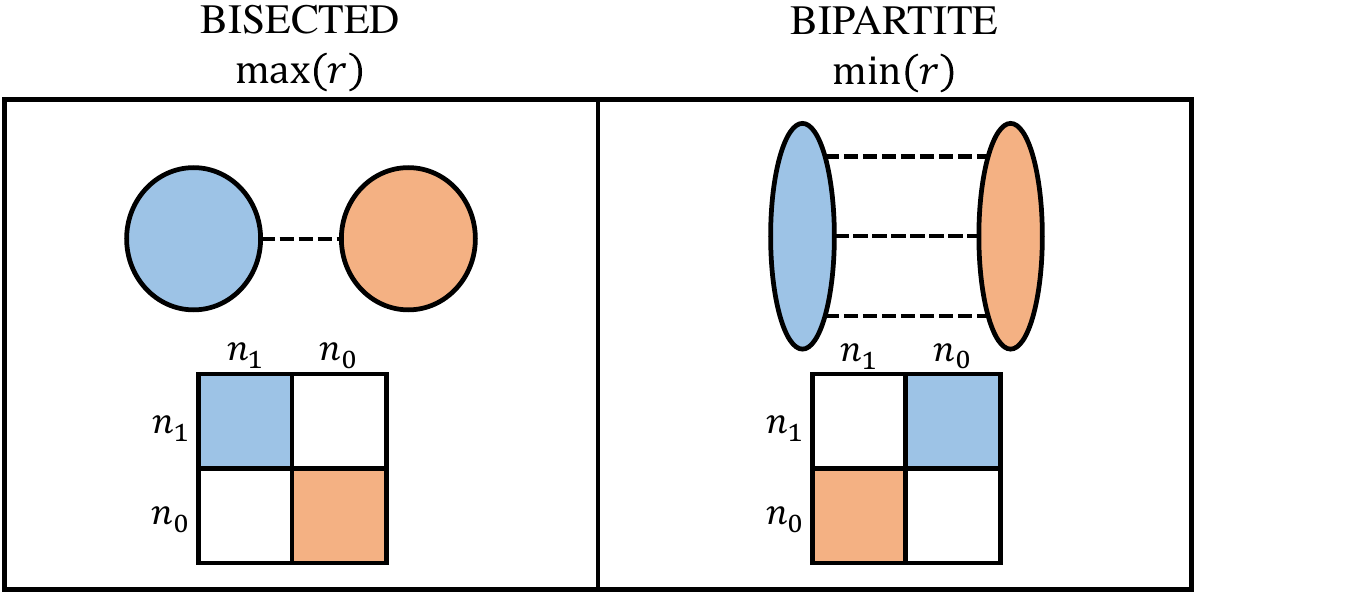}
   \caption{}
   \label{fig:dragratio3}
\end{subfigure}
\centering
\caption{
(a) Graphical representation of the assortativity range in the metadata-graph space, graph space and metadata space. 
The assortativity range in the graph space is represented differently as it may be either smaller or larger than the assortativity range in the metadata space.
(b) Graphical representation of block diagrams related to two different types of partition.
In the case of a bipartite-like configurations most of the links fall between different partitions (i.e., connect nodes with different metadata 
values) while in the case of a bisected-like configurations most of the links fall within each of the partitions (i.e., connect nodes with the same metadata values).
}
\label{alltogether}
\end{figure}

Figure~\ref{alltogether}(a) illustrates qualitatively the relationship of the bounds in each of these spaces. Both the metadata space and the graph space are subsets of the metadata-graph space. 
In what follows, we will demonstrate that the relationship between the bounds on the metadata-graph space and the bounds on the graph space is relatively straightforward. However, the relationship between the graph space and the metadata space in terms of assortativity is somewhat more nuanced. 
Depending on the topology and the node metadata vector, the assortativity range in the graph space can be either narrower or wider than the metadata space. 

Using both combinatorial and empirical methods, we demonstrate that these bounds can be substantially far from $-1$ and $1$ in each of these three spaces. 
We reinforce these  results by demonstrating that for some real-world networks the full reference range of assortativity is not attainable in any of these three spaces. Such evidence may provide some insights about the interpretation of common configurations as well as boundary ones that, without such a knowledge, would be misinterpreted by being considered less significant than they are.

\section{Assortative mixing of binary metadata}
\label{AEC}

The formula for Newman's assortativity, $r$, for binary node metadata is equivalent to Pearson's correlation coefficient for binary variables, also known as the $\phi$ coefficient~\cite{yule1912on}.  
Pearson's correlation coefficient is a fundamental statistic used to identify linear relationships between variables. However, despite the widespread use of Pearson's correlation coefficient, it is not without its limitations e.g., the ambiguity of interpreting specific coefficient values~\cite{anscombe1973graphs}. Given the relationship between Newman's assortativity and Pearson's correlation coefficient it is unsurprising that the assortativity coefficient also suffers from similar issues of interpretability~\cite{peel2018multiscale}.   
Here we consider a further issue that affects the interpretation of assortativity -- the extremal values $r \in \{-1,1\}$ are often unattainable. This particular issue is one that is inherited, in part, from the $\phi$ coefficient~\cite{Cureton1959}. However, in the case of network metadata, the effect is exacerbated by the dependency on the network structure. 

\subsection{The $\phi$ coefficient}
We begin by clarifying the relationship between the $\phi$ coefficient and Newman's assortativity, $r$. Since $\phi$ is a correlation coefficient, we can write the $\phi$ coefficient of binary variables $x$ and $y$ as:
\begin{equation}
 \phi = \frac{\mathbb{E}[x,y] - \mathbb{E}[x]\mathbb{E}[y]}{\sigma_{x}\sigma_{y}} \enspace ,
  \label{eq_pearson}
\end{equation}
 where $\sigma_{x}$ is the standard deviation of $x$.  For binary variables, the sample $\phi$ coefficient is based on the $2 \times 2$ contingency table: 
\begin{equation}
\textrm{P}(x,y) = 
 \begin{tabular}{c|cc|c}
& $y=0$ \, & \, $y=1$ & \\
 \hline
$x=0$ & $e_{00}$ & $e_{01}$ & $\quad a_0$ \\
$x=1$ & $e_{10}$ & $e_{11}$ & $\quad a_1$ \\
 \hline
 & $b_0$ & $b_1$ &
 \end{tabular}
 \qquad
 \textrm{where}
 \qquad a_i = \sum_j e_{ij} \quad b_j = \sum_i e_{ij}
 \enspace ,
 \label{eq_contingency}
\end{equation} 
and $e_{ij}$ is the proportion of pairs for which $x=i$ and $y=j$, and $\sum_{ij} e_{ij} = 1$. The $\phi$ coefficient~\cite{yule1912on} is stated as:
\begin{equation}
 \phi = \frac{e_{11} - a_{1}b_{1}}{\sqrt{a_{1}a_{0}b_{1}b_{0}}} \enspace .
 \label{eq_phi}
\end{equation}

Treating $e_{ij}$ as the joint probability distribution $\textrm{P}(x,y)$, the $\phi$-coefficient tells us the correlation of the variables $x$ and $y$ sampled from this distribution. In this setting the range of $\phi$ is bounded by the marginals~\cite{guilford1965minimal} (see Appendix~\ref{sec_phi_bounds}),
\begin{equation}
  \phi_{\min} = - \sqrt{\frac{a_0b_0}{a_1b_1}}
  \qquad 
  \phi_{\max} = \sqrt{\frac{a_0b_1}{a_1b_0}}  \enspace .
  \label{eq_phi_min_max}
\end{equation}

\subsection{Assortativity for binary node metadata}
Newman's assortativity for binary node metadata is calculated according to the same formula as Eq.~\eqref{eq_phi}. However, since we are considering the correlation of node metadata across network edges we no longer have two variables $x$ and $y$, we have a single binary variable $c$ representing the node metadata.
In undirected networks, $e_{ij}=e_{ji}$ and represents half the proportion of edges in the network that connect nodes with type $i$ to nodes with type $j$ (or the proportion of edges if $i = j$) and $a_i = b_i$, i.e.,
%
%
\begin{equation}
\textrm{P}(c_i,c_j) = 
 \begin{tabular}{c|cc|c}
& $c_i=0$ \, & \, $c_j=1$ & \\
 \hline
$c_i=0$ & $e_{00}$ & $e_{01}$ & $\quad a_0$ \\
$c_j=1$ & $e_{01}$ & $e_{11}$ & $\quad a_1$ \\
 \hline
 & $a_0$ & $a_1$ &
 \end{tabular}
 \qquad
 \textrm{where}
 \qquad a_i = \sum_j e_{ij} = \sum_i e_{ij}
 \enspace ,
 \label{eq_contingency_undir}
\end{equation}  
To make the connection between $\phi$ and Newman's assortativity explicit, first note that the numerator of Eq.~\eqref{eq_phi} can also be written as $e_{00} - a_{0}^2$ since $e_{11} - a_{1}^2 = e_{00} - a_{0}^2$. Now we can simplify the denominator of Eq.~\eqref{eq_phi},
\begin{align}
 \sqrt{a_{1}a_{0}b_{1}b_{0}} & = a_{1}a_{0} \notag\\
 & = a_1 (1 - a_1) \notag \\
 & = a_1 - a_1^2 \enspace . \notag
\end{align}

Then by making these substitutions and summing over categories (since we do not assume that $a_0 = a_1$) we recover Newman's assortativity~\cite{newman2003mixing}:
\begin{equation}
r = \frac{\sum_i e_{ii} - a_{i}^2}{1 - \sum_i a_{i}^2} \enspace .
\label{eq_newman_assort}
\end{equation}

The fact that the minimum value of assortativity $r_{\min}$ may be greater than $-1$ was previously indicated in~\cite{newman2003mixing} where a relatively conservative bound was derived by inspecting Eq.~\eqref{eq_newman_assort} and considering that the minimum occurs when all edges connect nodes with different metadata values, such that $\sum_i e_{ii}=0$. In this case, we see that the minimum assortativity is 
\begin{equation}
r_{\min} = \frac{-\sum_i a_{i}^2}{1 - \sum_i a_{i}^2} \enspace .
\label{eq_newman_min}
\end{equation}
However, this bound is not particularly informative in the case of binary metadata since it will always yield a value $\leq-1$. The reason is because it necessitates that a configuration where $\sum_i e_{ii}=0$ is possible. 
When the marginals of the metadata values are imbalanced, such that $a_0 \neq a_1$, then it is not possible for both $e_{00}=e_{11}=0$ to equal zero, since $e_{00}-a_0^2=e_{11}-a_1^2$. 
A more informative bound is that of the $\phi$-coefficient~\cite{guilford1965minimal},
\begin{equation}
 \phi_{\min} = - \frac{a_0}{a_1} \geq -1 \enspace ,
\label{eq_rmin_phimin}
\end{equation}
which we obtain from Eq.~\eqref{eq_phi_min_max} when $a_i=b_i$ and assuming that $a_0 \leq a_1$. 
When the marginals of the metadata values are balanced $a_{0}=a_1$ the metadata can potentially form a bipartite partition of the network such that the bound given in Eq.~\eqref{eq_rmin_phimin} is saturated. However, we will show that, depending on the space of configurations considered, it is not always possible to arrange the metadata to form a bipartite split. In these cases $r=-1$ will be unattainable. 
Regarding the maximum assortativity, we see in Eq.~\eqref{eq_phi_min_max} that $r_{\max} = \phi_{\max} = 1$ for an undirected network.

These bounds, however, only take into account the marginals $a_0$ and $a_1$, but as we see in Figure~\ref{correlation_assortativity} the underlying network presents extra constraints e.g., we might know that the network has a specific degree sequence. The model for correlation, in which we sample pairs from the joint distribution $\textrm{P}(x,y)$ of Eq.~\eqref{eq_contingency}, does not allow us to incorporate information about the network structure. This discrepancy suggests that this simple model is unsuitable and that modeling assortativity is a little more nuanced.

In order to include this information, we consider the model of assortativity originally proposed by Newman~\cite{newman2003mixing}, in which we sample from a joint distribution over node pairs $(i,j)$ in the adjacency matrix (such that existing edges are sampled uniformly at random) and look at the metadata values of the nodes linked by the edge. The probability of observing pairs of metadata values in this way is our joint distribution $\textrm{P}(c_i,c_j)$.
This model of assortativity naturally incorporates network properties such as the degree distribution. Ensuring a specific degree sequence (in expectation) amounts to fixing the row and column sums of the distribution over the adjacency matrix. Note that assuming this model of assortativity does not change how we calculate the assortativity coefficient since we aggregate counts by node metadata values to produce a contingency table [Eq.~\eqref{eq_contingency}] just as we would when calculating the $\phi$ coefficient. 
However, the additional structure that this model of assortativity affords limits the possible joint distributions $\textrm{P}(c_i,c_j)$ that we may consider and in turn further limits the attainable values of assortativity.

In what follows, it will become clearer to describe assortativity in terms of edge counts, rather than proportion of edges. To do so we consider that the $m$ edges are divided into three subsets such that $m = m_{11} + m_{10} + m_{00}$.  Then we make the simple substitution $m_{ij} = (2-\delta_{ij}) e_{ij}m$, where $\delta$ is the Kronecker delta. 
Consequently the assortativity of binary node metadata can be written as:
\begin{equation}
r = \frac{ (m_{00} + m_{11})m - (m_{00} + \frac{m_{10}}{2})^2 - (m_{11} + \frac{m_{10}}{2})^2}{m^2 - (m_{00} + \frac{m_{10}}{2})^2 - (m_{11} + \frac{m_{10}}{2})^2} \enspace ,
\label{binr}
\end{equation}
which can be simplified by first eliminating $m_{10}$ through the substitution $m_{10} = m - m_{00} - m_{11}$,
\begin{equation}
  r = \frac{2(m_{00} + m_{11})m - m^2 - (m_{00} - m_{11})^2}
  {m^2 - (m_{00} - m_{11})^2} \enspace ,
\end{equation}
and rearranging as:
\begin{equation}
  r = 1 - \frac{2m_{10}m}{m^2 - (m_{00} - m_{11})^2} \enspace .\label{binr_simp}
\end{equation}

The bounds on assortativity are directly related to the bounds on the edge counts. Specifically, we can consider two types of bipartition depending whether we wish to minimise or maximise the assortativity. Figure~\ref{alltogether}(b) illustrates this relationship. 
Maximising the assortativity corresponds to forming a minimum cut bisection of the network such that the majority of the edges connect nodes of the same type [Fig.~\ref{alltogether}(b)(\textit{left})]. The maximum value $r =1$ occurs when edges only occur between nodes with the same metadata values, i.e., $m_{00} + m_{11} = m$ and implies that the network is made up of multiple connected components, each containing only nodes with the same metadata value. 
Minimising the assortativity corresponds to finding a bipartite (or near bipartite) partition such that all (or most) of the edges connect a node $i$ that has metadata $c_i = 0$ to a node $j$ that has metadata $c_j=1$ [Fig.~\ref{alltogether}(b)(\textit{right})]. 
The minimum of $r=-1$ occurs if and only if $m_{00} + m_{11} = 0$. 

\section{Bounds on the edge counts using the degree sequence} 
\label{assortativity_b}
There are instances in which the bounds for assortativity, $-1 \leq r \leq 1$, can be attained. However, this is often not the case when certain properties of the network are fixed. In particular, the degree sequence, a specific set of edges and the way that node metadata values are assigned to specific nodes all play a role in limiting the range of attainable values of assortativity. 

Instrumental to exploring the effect of structural properties on the bounds of assortativity is the dependence of assortativity on the edge counts $m_{11}$, $m_{10}$ and $m_{00}$, as shown in Eq.~\eqref{binr_simp}. To demonstrate, we consider the largest of the aforementioned ensembles of graphs, the metadata-graph space (mgs), in which we preserve only the degree sequence of the observed graph and the relative proportions of observed metadata values.  
Within this space we can state bounds on the possible edge counts $m_{11}$, $m_{10}$ and $m_{00}$~\cite{cinelli2017structural, cinelli2019evaluating}. We denote the upper bounds with a superscript $u$ (e.g., $m_{11}^u$) and the lower bounds with a superscript $l$ (e.g., $m_{11}^l$).

In a graph with $n$ nodes that have a binary metadata assignment, we have $n_0$ nodes with metadata value $c_i = 0$ and $n_1$ nodes with metadata value $c_i = 1$.  We define bounds on the edge counts by partitioning the ordered degree sequence $D_G$ and using this partition of the degree sequence to consider the maximum and minimum edge counts that it imposes. 
For example, to determine the upper bound of the number of edges $m_{11}^u$ that connect pairs of nodes with metadata value 1 we should consider that the maximum value of $m_{11}$, given $n_0$ and $n_1$, occurs when $n_1$ nodes are arranged into a complete subgraph. If $D_G$ does not allow such a configuration, then the maximum value of $m_{11}$ occurs when $n_1$ nodes with the highest degree only connect to each other and not to any nodes with metadata $c_i=0$. Therefore we partition the degree sequence into a head $D_G^H(n_1)$, comprised of the $n_1$ highest degrees, and a tail $D_G^T(n_0)$, containing the $n_0$ lowest degrees, 
such that $D_G = D_G^H(n_1) \cup D_G^T(n_0)$.

\begin{figure}[!htb]
\begin{center}
\includegraphics[scale=.4, trim=0cm 3.5cm 0cm 5cm, clip=true]{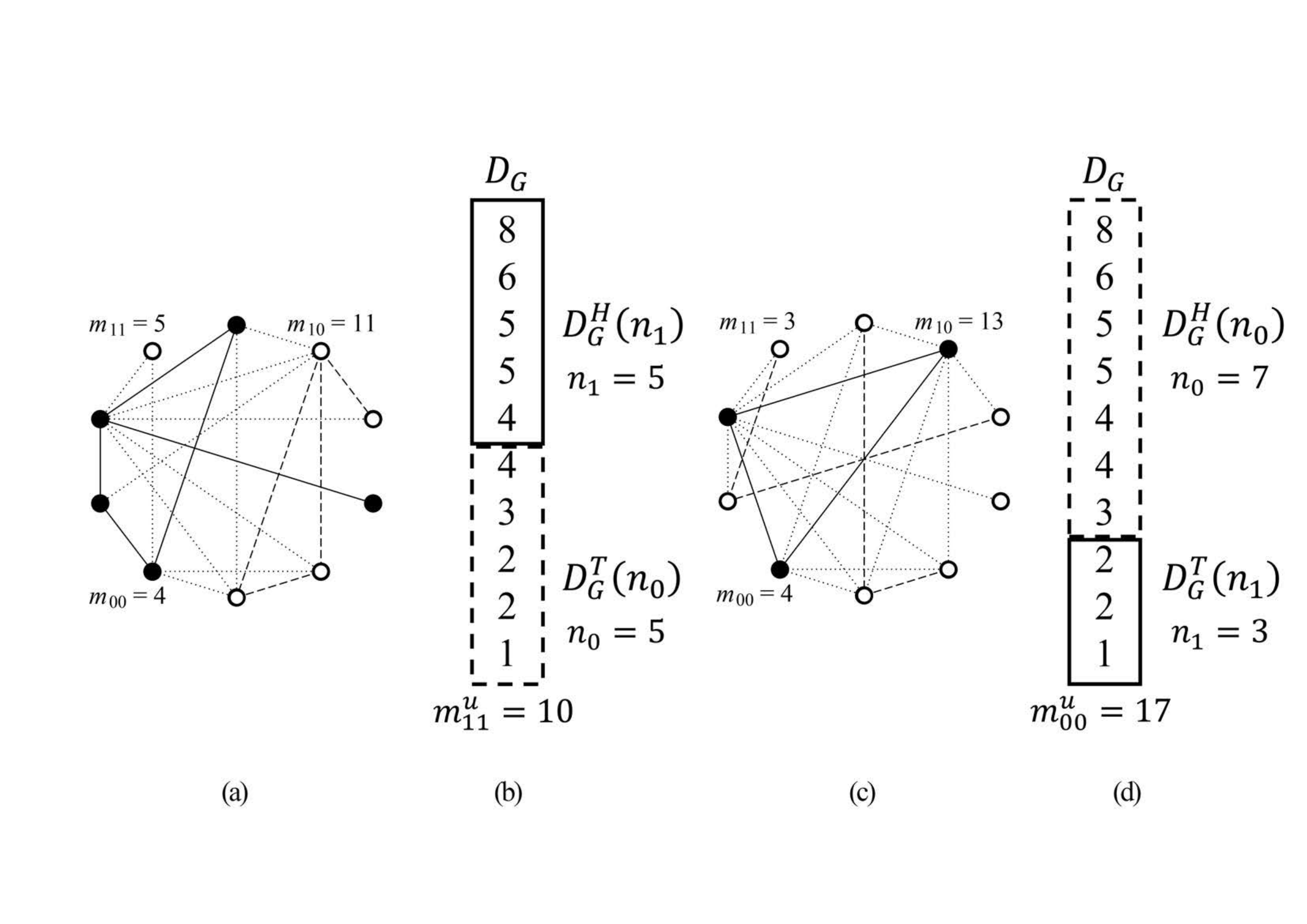}
\caption{Two different networks with the same degree sequence $D_G$, one with balanced proportions of metadata values, as displayed in panel (a), and the other with imbalanced proportions, as displayed in panel (c). 
The $m_{11}$ links between pairs of nodes with metadata $c_i =1 $ (in black) are represented by solid lines, the $m_{00}$ links between pairs of nodes with metadata  $c_i=0$ (in white) are represented by dashed lines, and the remaining $m_{10}$ links are represented by dotted lines. 
In each case we partition the degree sequence $D_G$ into a head $D_G^H$, containing the highest degrees, and a tail $D_G^T$, containing the lowest degrees, i.e., $D_G = D_G^H(n_1) \cup D_G^T(n_0)$, as shown in panel (b) (or $D_G = D_G^H(n_0) \cup D_G^T(n_1)$, as shown in panel (d)). Once we fix the degree sequence $D_G$, the bounds depend only on the proportion of featured nodes $n_1$ and $n_0$. Consequently, when $n_1=5$ (network in panel (a)) the bounds are $m_{11}^u=10$,  $m_{10}^u=20$, $m_{00}^u=10$, $m_{11}^l=0$, $m_{10}^l=4$ and $m_{00}^l=0$. When $n_1 = 3$ (network in panel (c)) the bounds are  $m_{11}^u=3$, $m_{10}^u=18$, $m_{00}^u=17$, $m_{11}^l=0$, $m_{10}^l=1$ and $m_{00}^l=1$.}
\label{NetandDs}
\end{center}
\end{figure}

Figure~\ref{NetandDs} shows a simple example of such a partition of the degree sequence with two different values of $n_1$. 
The maximum possible $m_{11}$ in any network that has degree sequence $D_G$ and $n_1$ has to be necessarily less than or equal to the number of links contained in the subgraph with degree sequence $D_G^H(n_1)$ (when we consider $D_G^H(n_1)$ the degree-sum of the $n_1$ elements of the network is maximized) or to the number of links contained in a clique of size $n_1$. 
The same reasoning is applicable when we consider $m_{00}$, whose upper bound can be computed by partitioning $D_G$ in a way such that $D_G = D_G^H(n_0) \cup D_G^T(n_1)$, as shown in panel (d) of Figure~\ref{NetandDs}.
Following a similar rationale of partitioning the ordered degree sequence, it is possible to either maximize or minimize the degree-sum of the two groups thus obtaining specific upper and lower bounds to the edge counts. Full details on the derivation of the upper and lower bounds ($m_{11}^u$, $m_{10}^u$, $m_{00}^u$) and ($m_{11}^l$, $m_{10}^l$ and $m_{00}^l$) are given in Appendix~\ref{EC}.

The obtained bounds require only the degree sequence and the proportion of metadata to be set and so are suitable for the metadata-graph space. However, they can be trivially extended to the graph space by considering the case of a fixed partition of the degree sequence as explained in Appendix~\ref{ECGS}. 
\section{Bounds on Binary Assortativity}
We now discuss bounds on assortativity relative to the limits imposed by the three spaces: the metadata-graph space (mgs), the graph space (gs) and the metadata space (ms). Throughout we will assume that $0 < n_1 < n$ to ensure that neither of the groups in the partition are empty.

\subsection{Bounds for the metadata-graph space}
The metadata-graph space contains all configurations of graph structures and node metadata assignments that have a specified degree sequence $D_{G}$ and given number of nodes of each type $\{n_0,n_1\}$. 
The bounds on the edge counts described in Section~\ref{AEC} depend only upon the specific degree sequence and the number of nodes of each type. We can therefore use these directly to define the bounds upon the metadata-graph space.

\subsubsection{Upper bound}
\label{assortativity_ub}
\begin{figure}[htbp]
\begin{center}
\includegraphics[scale=.75, trim=0cm 0cm 0cm 0cm, clip=true]{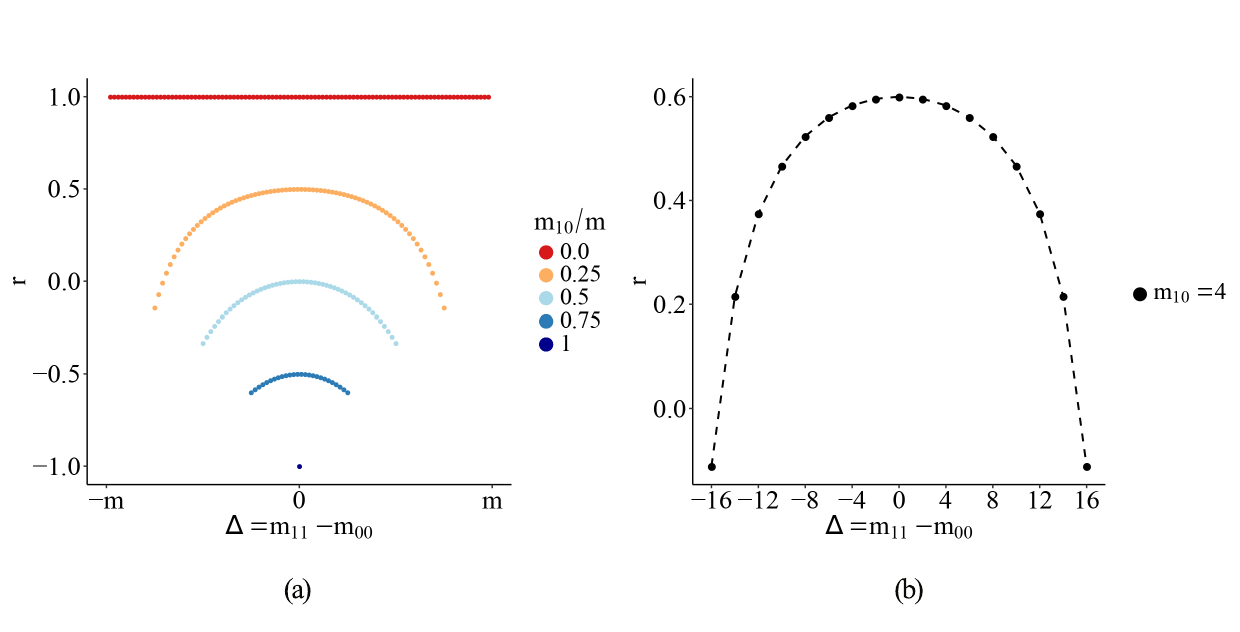}
\caption{Assortativity as a function of the difference in edge counts $m_{11} - m_{00}$, as computed from Eq.~\eqref{binr_simp}. Panel (a) displays the assortativity function $r$ for different values of $m_{10}$ as a proportion of $m$ while the difference between $m_{11}$ and $m_{00}$, represented by $\Delta=m_{11}-m_{00}$, varies. We notice that assortativity is maximized whenever $m_{11} = m_{00}$. Panel (b) refers to the degree sequence displayed in Fig.~\ref{NetandDs} when $n_1=5$. In this case we set $m_{10}=m_{10}^l=4$, which implies $m - m_{10} = m_{11} + m_{00} = 16$. Thus, we obtain the maximum value of assortativity when $m_{11}=m_{00}=(m - m_{10}^l)/{2}$ i.e., when $(m - m_{10}^l)/{2}= 8$.}
\label{UBDelta}
\end{center}
\end{figure}

The maximum value of $r_{\textrm{mgs}}$, for connected networks, occurs when as few edges as possible link nodes of different types. Therefore we define our upper bound $\rumgs$ by setting $m_{10} = m_{10}^l$. 
The maximum value of assortativity $r=1$ can only be attained if the graph can be partitioned into disconnected components that contain only a single type of node. This constraint implies that when the lower bound $m_{10}^l$ is greater than zero, then the maximum possible value of assortativity is less than 1. 
Setting $m_{10} = m_{10}^l$ implies that $m_{11} + m_{00} = m - m_{10}^l$ and we can write $r$ as:
\begin{equation}
r = 1 - \frac{2 m_{10}^l m}{m^2 - (m_{00} - m_{11})^2} \enspace .
\label{ub1}
\end{equation}
We can substitute $m_{00} = m - m_{10}^l - m_{11}$ into Eq.~\eqref{ub1} and write $r$ as a function of $m_{11}$,
\begin{equation}
r = 1 - \frac{2 m_{10}^l m}{m^2 - (m - m_{10}^l - 2m_{11})^2} \enspace .
\label{ub2}
\end{equation}
In order to obtain the value of $m_{11}$ that maximizes $r$ we can solve the following equation:
\begin{equation}
\label{supp_ub3}
\frac{\partial r}{\partial m_{11}} = 0 \enspace ,
\end{equation}
which gives us $m_{11} = (m-m_{10}^l)/{2}$ and implies that $m_{00} = (m - m_{10}^l)/{2} = m_{11}$ and therefore: 
\begin{equation}
\label{maxr}
\rumgs = 1 - \frac{2m_{10}^l}{m} \enspace .
\end{equation}
Figure~\ref{UBDelta} graphically confirms that $r$ is maximized when $m_{00}=m_{11}$ because assortativity, with fixed $m_{10}$, is a concave function of $m_{11}$. 

\subsubsection{Lower bound}
\label{assortativity_lb}
The minimum value of $r_{\textrm{mgs}}$, for connected networks, occurs when the partition of the node metadata forms a bipartite split of the graph. 
When $m_{11}^l + m_{00}^l > 0$ for a given degree sequence $D_{G}$ and group sizes $n_0$ and $n_1$ it means that a certain number of intra-group links exist and that it is not possible to form a bipartite partition $G_{n_1, n_0}$ of the graph.
As introduced in Sec.~\ref{AEC}, Newman provides the following lower bound to assortativity~\cite{newman2003mixing}:
\begin{equation}
r_{\min} = -\frac{\sum_i a_i^2}{1 - \sum_i a_i^2} \enspace . \label{eq_rmin_newman}
\end{equation}

Such a lower bound assumes the existence of a bipartite split of the nodes and that the sum of the proportion of intra-partition links is zero, i.e., $\sum_i e_{ii}=0$. Written in terms of edge counts for binary metadata we have $m_{11} + m_{00} =0$, which $m_{10} = m$ (since $m_{11}+m_{00}+m_{10}=m$). By substituting these quantities into Eq.~\eqref{eq_rmin_newman} we obtain:
\begin{equation}
r_{\min} = -\frac{2(\frac{m_{10}}{2m})^2}{1 -2(\frac{m_{10}}{2m})^2} = -1 \enspace .
\end{equation}
Following Newman's reasoning but this time accounting for the bounds on the edge counts:
$m_{11} \geq m_{11}^l \geq 0$, $m_{00} \geq m_{00}^l \geq 0$ and $m_{10} \leq m_{10}^u \leq m$,
we obtain the lower bound $\rlmgs$ by considering a realization as close to a bipartite split of the graph (i.e. that with the highest $m_{10}$) as possible, which presents us with two options. 
The first option consists in setting $m_{11} = m_{11}^l$ and $m_{00} = m_{00}^l$, which implies $m - m_{11}^l - m_{00}^l= m_{10}$. Through a simple substitution of such quantities we obtain:
\begin{equation}
m_{ii}^l \textrm{ case} : \rlmgs = 1 - \frac{2m(m - m_{00}^l - m_{11}^l)}{m^2-(m_{00}^l-m_{11}^l)^2} \enspace .
\end{equation}

The second option is to use the upper bound of the edge count $m_{10}$ by setting $m_{10} = m_{10}^u$. 
When $m_{10} = m_{10}^u$ then $m_{11} + m_{00} = m - m_{10}^u$. For a fixed value of $m_{10}$, binary assortativity is a concave function and $m_{00}^l$ can be different from $m_{11}^l$. 
When $m_{11} + m_{00} > 0$ the minimum assortativity can be obtained when the absolute difference $|\Delta| = |m_{11} - m_{00}|$ is maximized (as shown in Figure~\ref{UBDelta}).
We then have two further options to determine the lower bound. 
In the first case we set $m_{10}=m_{10}^u$, $m_{11}=m_{11}^l$ and if $m_{10}^u+m_{11}^l \leq m$ then $m_{00} = m-m_{10}^u-m_{11}^l$ which we refer to as the $\Delta_{\min}$ case. Thus:
 \begin{equation}
 \label{min2}
 \Delta_{\min} \textrm{ case}: \rlmgs = 1 - \frac{2 m_{10}^u m}{m^2 - (m-m_{10}^u-2m_{11}^l)^2} \enspace .
 \end{equation} 

In the second case we set $m_{10}=m_{10}^u$, $m_{00}=m_{00}^l$ and if $m_{10}^u+m_{11}^l \leq m$ then $m_{11} = m-m_{10}^u-m_{00}^l$ which we refer to as the $\Delta_{\max}$ case. Thus:
 \begin{equation}
 \label{min3}
 \Delta_{\max} \textrm{ case}:\rlmgs = 1 - \frac{2 m_{10}^u m}{m^2 - (m_{10}^u+2m_{00}^l-m)^2}
 \end{equation}
In summary, we have three possible cases for determining the lower bound:
 \begin{equation}
 \label{min_assort}
\{m_{00}, \; m_{11}, \; m_{10} \} = 
 \begin{cases}
  m_{ii}^l &: \left\lbrace m_{00}^l, \; m_{11}^l, \; m - m_{11}^l - m_{00}^l \right\rbrace 
    \\
  \Delta_{\max}&: \left\lbrace m-m_{10}^u-m_{11}^l, \; m_{11}^l, \;   m_{10}^u \right\rbrace
  \\
  \Delta_{\max}&: \left\lbrace m_{00}^l, \; m-m_{10}^u-m_{00}^l, \;   m_{10}^u \right\rbrace
 \end{cases} \enspace ,
\end{equation}
which we can substitute into Eq.~\eqref{binr} to obtain a lower bound $\rlmgs$ on the minimum value of $r$.

\subsection{Bounds for the graph space}
Similar to the metadata-graph space, the graph space considers all configurations of graph structures that have a specified degree sequence. The difference is that in the graph space the assignment of metadata values to nodes is fixed and so the degree-metadata correlation is also fixed. 
We can bound the assortativity of the graph space using the same rationale as the metadata-graph space. In fact, we can use the same equations given in the previous subsection by replacing the bounds on the edge counts for the metadata-graph space with those of the graph space, which are given in Section \ref{ECGS}.

The range of assortativity in the metadata-graph space is at least as large as the range in the graph space since the former has an extra degree of freedom by allowing all the possible arrangements of the metadata over the network nodes. As such, the following relations hold:
\begin{alignat}{2}
\rugs & \leq \rumgs \leq r_{\max} && \\
r_{\min} & \geq \rlmgs  \geq \rlgs && \enspace .
\end{alignat}

The difference in the assortativity bounds for the metadata-graph space and the graph-space vary according to $n_1$. The number of nodes $n_1$ controls how many distinct metadata assignments there can be per graph configuration. The ranges are equal when $n_1=\{0, n\}$, because there can be only a single metadata assignment for every graph configuration, and the difference in the ranges is maximized when $n_1 = n/{2}$. 

\subsection{Bounds for the metadata space}
Unlike the metadata-graph space and the graph space, in the metadata space we cannot provide any tighter theoretical bound for assortativity using only the degree sequence. Indeed, in order to obtain a tight bound in the metadata space we may need to exploit higher order statistics, such as the number of triangles or the diameter of the network.
Therefore we must resort to a complete enumeration, when feasible, or to a heuristic algorithm, based on a Monte Carlo exploration of the metadata space (one such algorithm is described in Section \ref{switch}).  
As in the previous section, we can clearly determine that the range of assortativity in the metadata-graph space is at least as large as the metadata space, because for each unique assignment of metadata values to node degrees, the metadata-graph space contains all possible graph configurations with the given degree sequence. Therefore,
\begin{alignat}{2}
\rums & \leq \rumgs \leq r_{\max} && \\
r_{\min} & \geq \rlmgs  \geq \rlms && \enspace ,
\end{alignat}
where $\rums$ and the $\rlms$ are the upper bound and the lower bound to the metadata space computed algorithmically.
We cannot guarantee, however, any relationship between the metadata space and the graph space since they are constrained by different elements. The former is constrained by the topology and by the proportion of metadata values while the latter by the degree sequence and by the assignment of metadata to specific nodes. 
Instead of a combinatorial bound in the metadata space, we demonstrate the bounds empirically on a synthetic network. 
Using the network in Figure~\ref{NetandDs} (top left network), we look at all the possible permutations of the metadata in order to compute the distribution of assortativity values.

\begin{figure}[htbp]
\begin{center}
 \begin{subfigure}[b]{0.45\textwidth}
   \centering
   \includegraphics[trim=0cm 0cm 0cm 0cm, clip=true, totalheight=0.35\textheight]{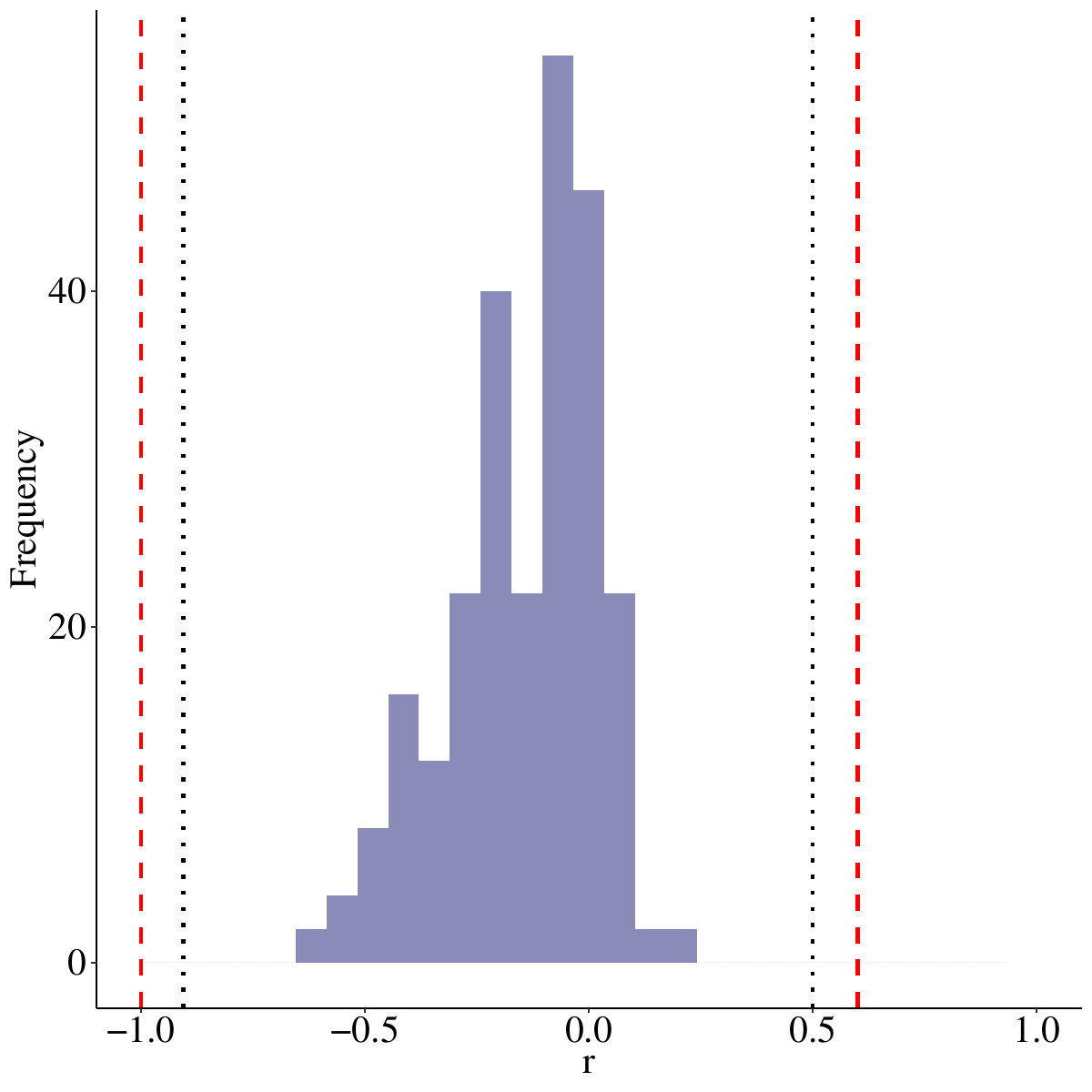}
   \caption{}
 \end{subfigure}
 \ \hspace{5mm} \hspace{5mm} 
 \begin{subfigure}[b]{0.45\textwidth}
  \centering
   \includegraphics[trim=0cm 0cm 0cm 0cm, clip=true, totalheight=0.35\textheight]{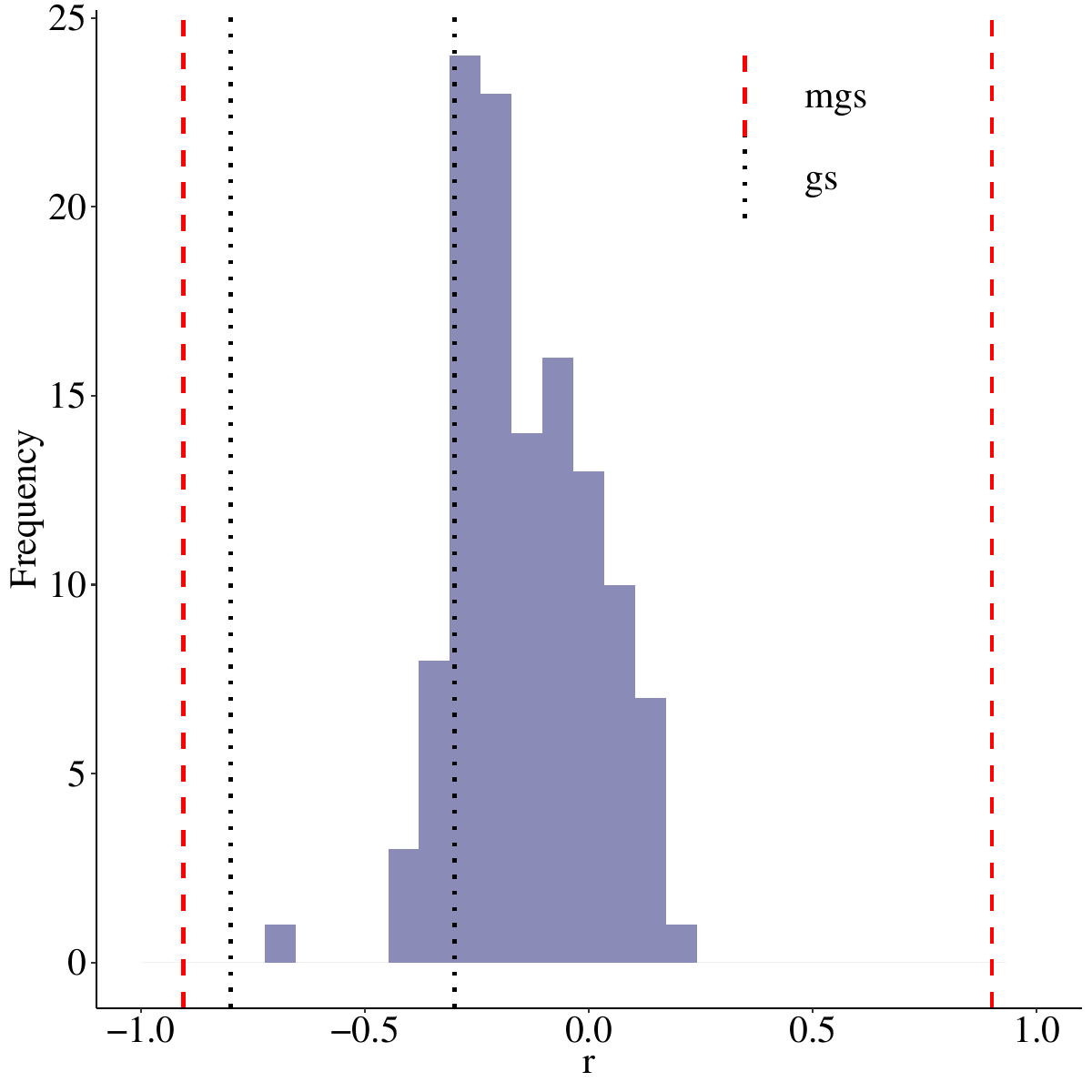}
   \caption{}
 \end{subfigure}
\end{center}
 \caption{Ranges of assortativity for the network shown in panel (a) of Fig.~\ref{NetandDs}. The plot on panel (a) has $n_1 = 5$ and the one on panel (b) has $n_1 = 3$. 
The histograms show the distribution of assortativity values in the metadata space (complete enumeration of all permutations of metadata assignments). The dashed lines indicate the bounds $\rlmgs$ and $\rumgs$ of the assortativity range in the metadata-graph space. The dotted lines indicate the bounds $\rlgs$ and $\rugs$ in the graph space. When $n_1 = 5$ the values are $\rlmgs = -1$, $\rlgs = -0.905$, $\rlms = -0.6$, $\rums = 0.2$, $\rugs = 0.5$, $\rumgs = 0.6$.
When $n_1 = 3$ these values are $\rlmgs = -0.905$, $\rlgs = -0.8$, 
 $\rlms = -0.704$, $\rugs = -0.3$, $\rums = 0.2$, 
 $\rumgs = 0.9$.
}
 \label{hist5and3n_1}
\end{figure}

Figure~\ref{hist5and3n_1} shows the distribution of assortativity values over a complete enumeration of metadata assignments for the network in Figure~\ref{NetandDs} (top left). The histogram on the left shows the distribution for $n_1=5$, while the one of the right shows the distribution for $n_1=3$.  Here we can clearly observe the minimum and maximum values of the assortativity coefficient attainable in the metadata space. 

\section{Experiments on real networks}
\label{exp}
In this section we investigate assortative mixing for binary metadata in real-world networks\footnote{Code is available online:\url{https://github.com/cinhelli/Network-constraints-on-the-mixing-patterns-of-binary-node-metadata}}. Examples of binary metadata of network nodes can be found in a wide array of contexts, including: the functional categories of proteins in protein-protein interaction networks~\cite{park2007distribution}, the hydrophobic/hydrophilic nature of proteins in protein contact networks~\cite{di2012protein}, and the use of a specific service in telecommunication networks~\cite{park2007distribution}. 
Here we will focus on another natural case study on binary node metadata, which is gender assortativity in social networks of animals~\cite{lusseau2004identifying} and humans~\cite{mcpherson2001birds}. 
The investigation of gender assortativity is interesting for a number of practical reasons related to human behaviour and the adoption of specific habits~\cite{laniado2016gender, centola2011anexperimental, cunningham2012isthereevidence, 
rosenquist2010spread}. Moreover, better understanding of the mixing patterns and preferences in social networks plays an important role in predicting missing metadata such as gender~\cite{altenburger2017bias}. 

Here we investigate gender assortativity in two colleges, Smith and Wellesley, extracted from the Facebook~100 dataset~\cite{traud2012social} (see Appendix~\ref{DSDescr} for dataset description) containing social network snapshots with a heavy-tailed distribution. Additional analytical results for scale-free networks are given in Appendix~\ref{BSF}.
Smith displays a gender assortativity of $r = 0.02$ that is positive but close to 0 (i.e., close to a random distribution of links relative to the node metadata), while in Wellesley we see a value of $r = 0.24$ that should indicate a distinctive pattern of assortativity by gender. 

\begin{figure}[htbp]
\begin{center}
 \begin{subfigure}[b]{0.45\textwidth}
   \centering
   \includegraphics[trim=0cm 0cm 0cm 0cm, clip=true, totalheight=0.35\textheight]{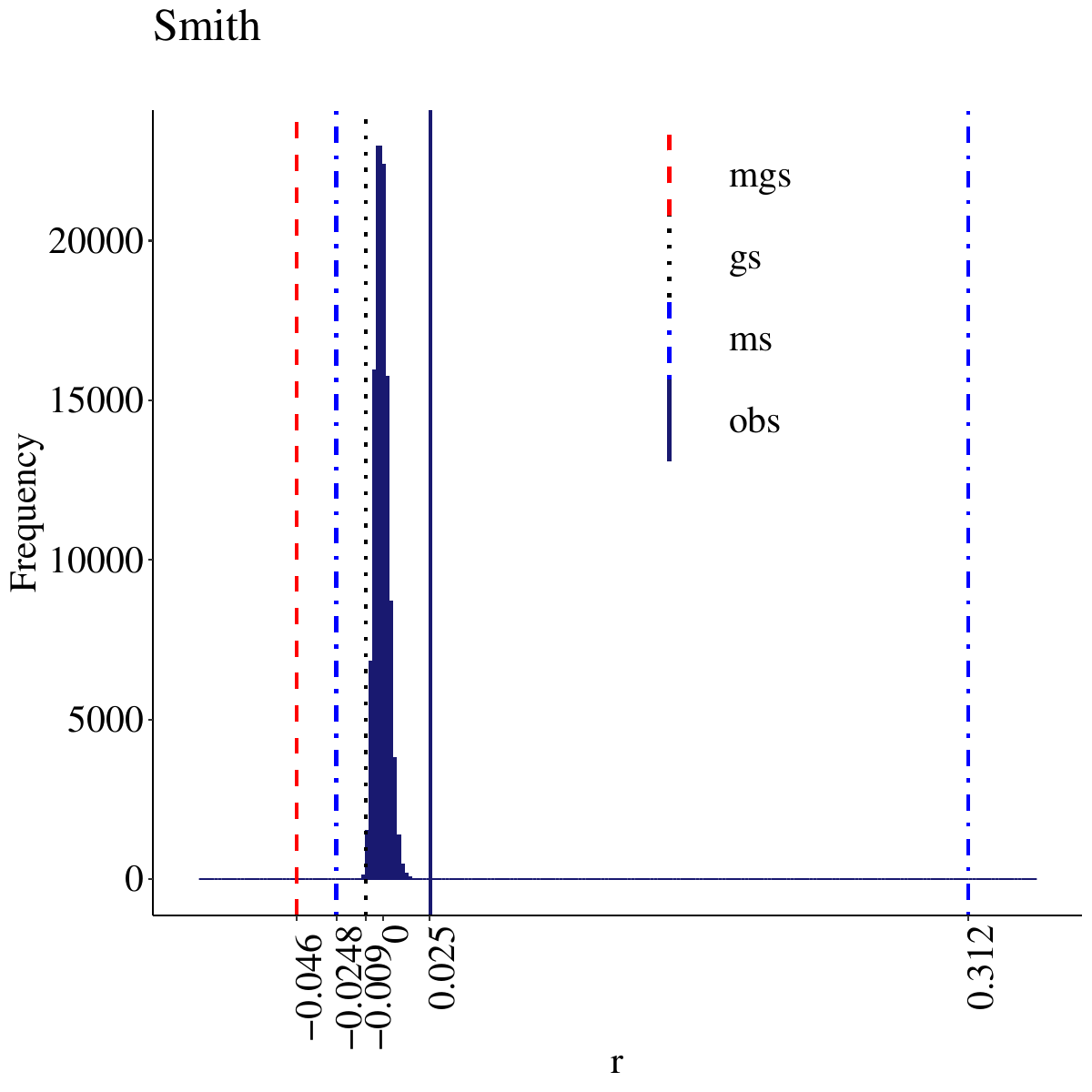}
   \caption{}
 \end{subfigure}
 \ \hspace{5mm} \hspace{5mm} 
 \begin{subfigure}[b]{0.45\textwidth}
  \centering
   \includegraphics[trim=0cm 0cm 0cm 0cm, clip=true, totalheight=0.35\textheight]{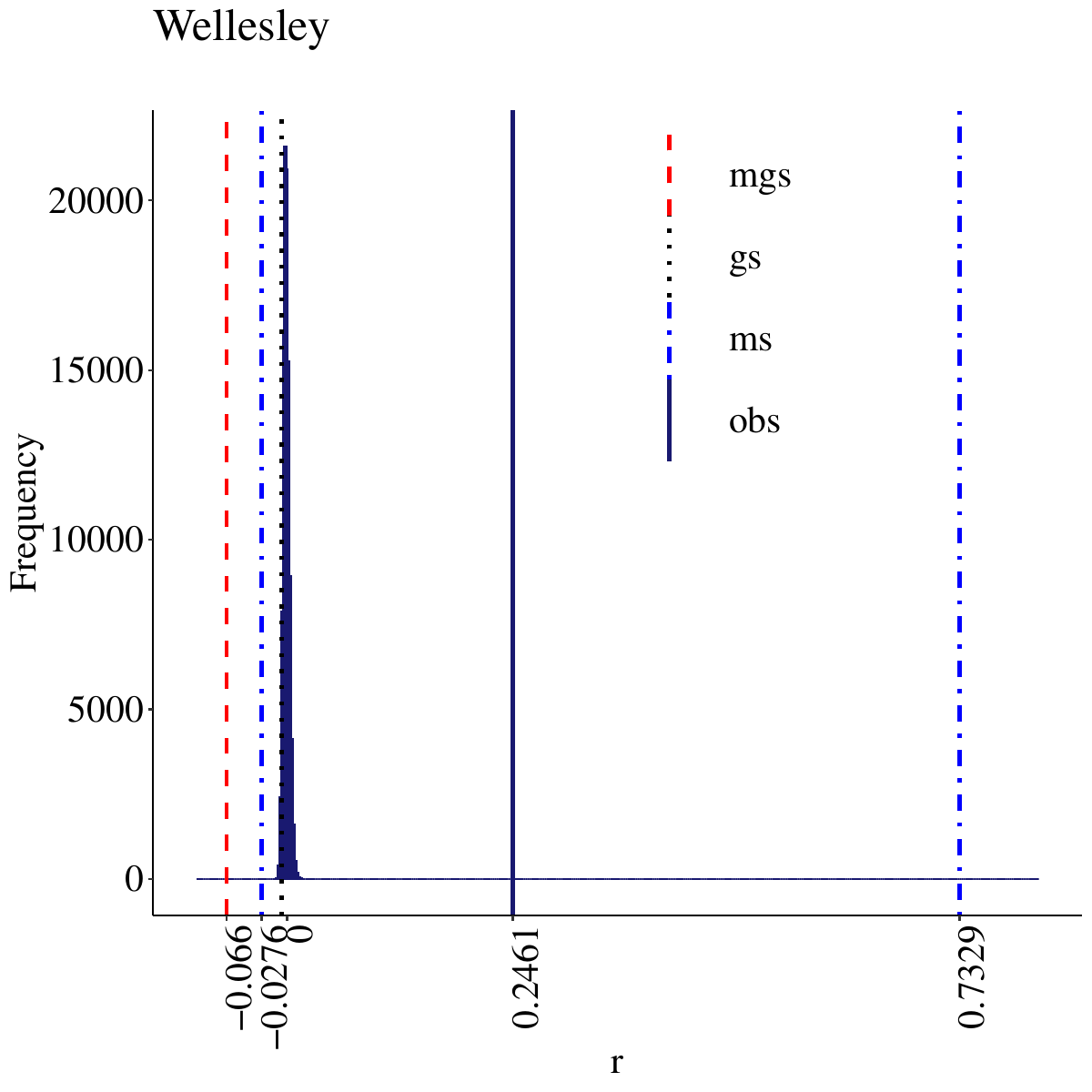}
   \caption{}
 \end{subfigure}
\end{center}
\caption{Assortativity bounds for the colleges Smith -- panel (a) -- and Wellesley -- panel (b) -- of the largest connected component after eliminating nodes with missing gender metadata. Smith college has $n =  2625$, $m = 77259$, $n_{\textrm{female}} = n_0 = 2596$, $n_{\textrm{male}} = n_1 = 29$. Smith has a gender assortativity value $r = 0.025$ (solid line) with $m_{11}=25$ and $m_{10}=1404$. The upper bounds to gs and mgs for Smith college are not reported in the panel and they are $\rugs = 0.976$ and $\rumgs = 1$. 
Wellesley college has $n =2689$, $m=78853$, $n_{females} = n_0 =2653$, $n_{males} = n_1 = 36$. Wellesley has a gender assortativity value $r = 0.246$ (solid line) with $m_{11}=122$ and $m_{10}=729$. The upper bounds to gs and mgs for Wellesley college are not reported in the panel and they are $\rugs = 0.995$ and $\rumgs = 1$.} 
 \label{Smith_Figure}
\end{figure}

Figure~\ref{Smith_Figure} displays the assortativity bounds for the metadata-graph and graph spaces for the Smith -- panel (a) -- and Wellesley -- panel (b) -- social networks. To evaluate the metadata space, Figure~\ref{Smith_Figure} also shows histograms of assortativity values for $10^5$ permutations of the binary metadata vector $\mathbf{c}$ for each network and the extremal values obtained from minimising and maximising the assortativity using an optimisation heuristic, described in Appendix~\ref{switch}. We immediately observe that the disassortativity of both networks is bounded away from -1, in all three spaces, such that the network cannot be very disassortative and that there are relatively few configurations of the network and metadata that allow disassortative mixing. This effect is partially due to the huge gender imbalance as both colleges are female only and so there are relatively few males (staff and graduate students) in the network. 

Comparing the ranges of assortativity in the metadata space and graph space, 
we observe that: $\rlms \leq \rlgs$ and $\rums \leq \rugs$ for both these networks. So, the metadata space allows more disassortative mixing than the graph space does. However, we should consider the latter with caution since $\rums$ is computed via a heuristic (thus it is a lower bound to the actual maximum) while $\rugs$ is an upper bound. Therefore, given the fact that $\rums$ and $\rugs$ have very close values it indicates a very high similarity in terms of upper bound of assortativity in graph and metadata spaces. 

With regard to the metadata space, we see that for both networks the observed assortativity value is higher than the assortativity of random permutations. We can interpret such a result as a test of statistical significance~\cite{peel2017ground}. So even though the assortativity is relatively low (particularly for the Smith network) we can still conclude that the assortativity is significantly higher than a random partition of the network ($p<10^{-5}$). 

In order to complement the previous analysis, we also consider a smaller but much denser network, the Wolf Dominance network~\cite{van1987dominance} (see Appendix~\ref{DSDescr}), over which we evaluate gender assortativity. 
In this smaller network it is possible to evaluate the metadata space via a complete enumeration of the possible metadata permutations and so we can compute the actual values of $\rlms$ and $\rums$ to compare against the combinatorial bounds of the graph space and metadata-graph space. 

Interestingly in this case, the upper bounds on assortativity in all three spaces are very close to zero and in the metadata space it is not possible to observe positive assortativity. Furthermore, the mean value of assortativity over the metadata space is not zero, as we can see from the histogram centred at ~-0.06. This observation seems contrary to our expectation that assortativity of random partitions should be centred around zero. This result resembles that of Ref.~\cite{fosdick2016configuring} in which it was observed that under certain conditions the expected value of assortativity in the \textit{graph space} is not equal to zero. 
Here we make a similar observation, that the expected value of assortativity may be different from zero, in the metadata space. Therefore, following a similar argument, we may conclude that in cases such as these an adjustment to the expected value may be necessary.
\begin{figure}[htbp]
   \centering
   \includegraphics[trim=0cm 0cm 0cm 0cm, clip=true, totalheight=0.34\textheight]{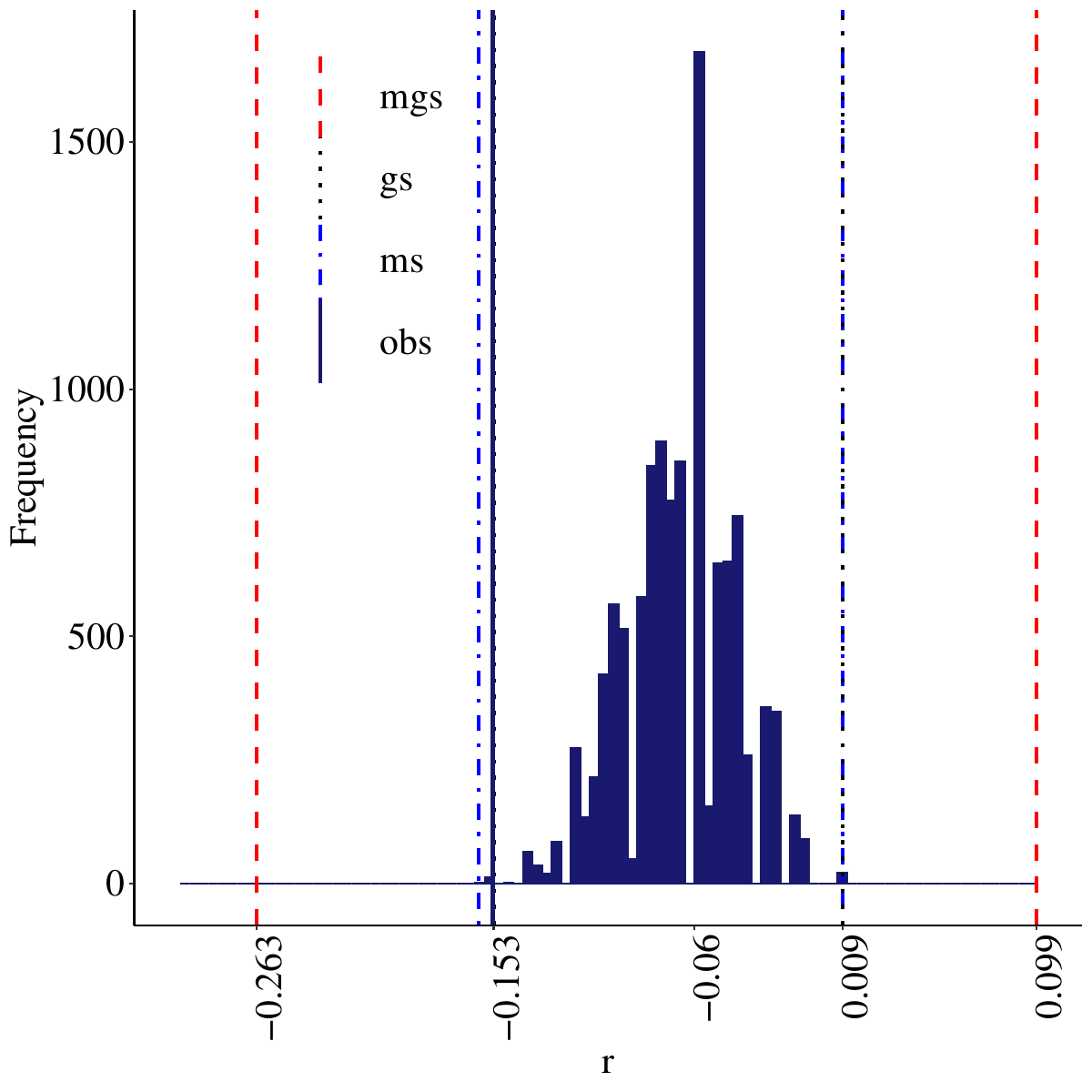}
 \caption{Assortativity bounds for the Wolf Dominance network ($n=16$, $m=111$, $n_{\textrm{female}} = n_0 = 7$, $n_{\textrm{male}} = n_1 = 9$).
The network has a gender assortativity value $r = - 0.153$ (solid line) that occurs in correspondence with $m_{11}=31$ and $m_{10}=63$. 
The dashed lines, obtained with the combinatorial bounds, occur in correspondence of $\rlmgs = -0.263$ and $\rumgs = 0.099$. The dot-dashed lines occur in correspondence of the values $\rlms = -0.16$ and $\rums = 0.009$ which are obtained via complete enumeration. The bounds to the graph space are represented by dotted lines at $\rlgs=-0.153$ and $\rugs=0.009$.}
\label{grid_Wolves}
\end{figure}

\section{Discussion}
The assortativity coefficient $r$ is generally assumed to range between $-1$ and $1$. 
In this paper we provide evidence that this widely used measure is often bounded within a much narrower range, given even partial knowledge of the metadata distribution or the network topology. It is important to be aware of these bounds when interpreting the values of assortativity. For instance, consider the example of the Wolf Dominance network in Figure~\ref{grid_Wolves}. The observed value of assortativity is $r=-0.153$. If we compare this value to the full interval $[-1,1]$, we might conclude that the network is mildly disassortative. However, we have established that all networks sharing the same degree sequence (i.e., in the same graph space) and the same proportion of male/female nodes as the original network will display assortativities bounded by the range $-0.153$ and $0.009$. In this light, we must conclude that the observed metadata are remarkably disassortative, in fact as disassortative as can be, given these constraints. Our work can, in this way, be used as tool that allows us to explore the significance of assortativity in a given network.

These findings will have impact on how we interpret assortativity. For instance in the study of the integration of minorities over time or across different networks. Taking into account the range of attainable values provides us with a clearer view on mixing patterns between such groups.
In summary, the range of values attainable by the assortativity coefficient can vary substantially depending on the constraints imposed by the network structure and the distribution of node metadata. Here we considered three particular settings:
\begin{itemize}
\item the \textit{metadata-graph space}: the range of assortativity values over the ensemble of configurations with a given degree sequence $D_G$ and number of nodes of each type $n_0,n_1$. Here we provide a combinatorial lower bound $\rlmgs$ and a combinatorial upper bound $\rumgs$.
\item the \textit{graph space}:  the range of assortativity values over the ensemble of configurations with a given degree sequence $D_G$ and a specific assignment of metadata to nodes. Here we provide a combinatorial lower bound $\rlgs$ and a combinatorial upper bound $\rugs$. 
\item the \textit{metadata space}: the range of assortativity values over the ensemble of permutations of the metadata values (preserving the counts $n_0, n_1$) on the specific topology of an observed graph. Here we propose the use of a heuristic to estimate the upper bound $\rums$ and lower bound $\rlms$.
\end{itemize}

The choice of space should depend upon the specific problem at hand and relate to the specific assumptions we wish to make about the graph structure and metadata assignment. For instance, when investigating metadata such gender in a social network we might consider the metadata and popularity of the nodes (i.e., their degrees) to be fixed and so the graph space might be most appropriate. Alternatively, in a road network in which the metadata indicates either presence or absence of road signals, occurrences of traffic jams or the locations of accident hotspots, we might want to consider the metadata space as the structure of the graph is fixed. 

Although we have focused on binary metadata, the issue of attaining the extremal values $\{-1,1\}$ of assortativity is still present for any categorical-valued metadata~\cite{cohen1960coefficient}. 
Taken altogether we can conclude that these constraints  present questions about the interpretability of network assortativity, especially when comparing across networks~\cite{traud2012social, jadidi2017gender, ugander2011anatomy, jacobs2015assembling}, as also noted in the case of degree assortativity~\cite{dorogovtsev2010zero}. As a potential solution, we might consider normalizing assortativity according to the bounds of the  space most relevant to our given problem. For instance, if we consider the degree and metadata value of a node to be fixed, then an appropriate normalization might be:
\begin{equation}
r^{\textrm{gs}} = 
\begin{cases}
  \frac{r}{\rugs} & \textrm{if $r$ is positive} \\
  \frac{r}{\rlgs} & \textrm{otherwise} \enspace .
\end{cases}
\end{equation}
Such a normalization has previously been suggested for related measures such as the $\phi$ coefficient and Cohen's~$\kappa$~\cite{Cureton1959,cohen1960coefficient}. It also follows the rationale that assortativity is a normalised version of modularity $Q$ (i.e., $r = Q/Q_{\max}$~\cite{newman2010networks} where $Q_{\max} = 1 - \sum_i a_{i}^2$). Alternatively we may consider comparing the observed assortativity with the distribution of assortativity values in the relevant ensemble, e.g., using assortativity as a test statistic in a one-sided hypothesis test to assess statistical significance~\cite{bianconi2009assessing,peel2017ground}. Additional investigations could expand the research outlined in Appendix~\ref{BSF} on the interplay between asymptotic properties of the degree distribution and the attainable range of assortativity in the case of categorical~\cite{cinelli2017structural} and scalar~\cite{yang2017lower} features. However, we leave the exploration of these ideas for future work.

Another avenue for future work would be to consider how a given ensemble constrains other network measures such as Freeman's segregation~\cite{freeman1978segregation}, which is limited by the edge count $m_{10}$ (see note in Appendix~\ref{segregation}), and the clustering coefficient, which is closely related to assortativity of scalar features such as degree~\cite{estrada2011combinatorial}.

\section*{Acknowledgements}
The authors thank Samin Aref, Alexandre Bovet, Andrea Cristofaro, Ernesto Estrada, Fariba Karimi, Dan Larremore, Giacomo Livan, Tiago Peixoto, Leo Pen, Martin Rosvall, Michael Schaub, Sofia Teixeira and Jean-Gabriel Young for helpful conversations. This work was supported in part by the F.R.S-FNRS Grant No.~1.B.336.18F [LP] and the Flagship European Research Area Network (FLAG-ERA) Joint Transnational Call ``FuturICT 2.0'' [JCD].


\begin{thebibliography}{36}%
\makeatletter
\providecommand \@ifxundefined [1]{%
 \@ifx{#1\undefined}
}%
\providecommand \@ifnum [1]{%
 \ifnum #1\expandafter \@firstoftwo
 \else \expandafter \@secondoftwo
 \fi
}%
\providecommand \@ifx [1]{%
 \ifx #1\expandafter \@firstoftwo
 \else \expandafter \@secondoftwo
 \fi
}%
\providecommand \natexlab [1]{#1}%
\providecommand \enquote  [1]{``#1''}%
\providecommand \bibnamefont  [1]{#1}%
\providecommand \bibfnamefont [1]{#1}%
\providecommand \citenamefont [1]{#1}%
\providecommand \href@noop [0]{\@secondoftwo}%
\providecommand \href [0]{\begingroup \@sanitize@url \@href}%
\providecommand \@href[1]{\@@startlink{#1}\@@href}%
\providecommand \@@href[1]{\endgroup#1\@@endlink}%
\providecommand \@sanitize@url [0]{\catcode `\\12\catcode `\$12\catcode
  `\&12\catcode `\#12\catcode `\^12\catcode `\_12\catcode `\%12\relax}%
\providecommand \@@startlink[1]{}%
\providecommand \@@endlink[0]{}%
\providecommand \url  [0]{\begingroup\@sanitize@url \@url }%
\providecommand \@url [1]{\endgroup\@href {#1}{\urlprefix }}%
\providecommand \urlprefix  [0]{URL }%
\providecommand \Eprint [0]{\href }%
\providecommand \doibase [0]{http://dx.doi.org/}%
\providecommand \selectlanguage [0]{\@gobble}%
\providecommand \bibinfo  [0]{\@secondoftwo}%
\providecommand \bibfield  [0]{\@secondoftwo}%
\providecommand \translation [1]{[#1]}%
\providecommand \BibitemOpen [0]{}%
\providecommand \bibitemStop [0]{}%
\providecommand \bibitemNoStop [0]{.\EOS\space}%
\providecommand \EOS [0]{\spacefactor3000\relax}%
\providecommand \BibitemShut  [1]{\csname bibitem#1\endcsname}%
\let\auto@bib@innerbib\@empty
\bibitem [{\citenamefont {Newman}(2003)}]{newman2003mixing}%
  \BibitemOpen
  \bibfield  {author} {\bibinfo {author} {\bibfnamefont {Mark~EJ}\ \bibnamefont
  {Newman}},\ }\bibfield  {title} {\enquote {\bibinfo {title} {Mixing patterns
  in networks},}\ }\href@noop {} {\bibfield  {journal} {\bibinfo  {journal}
  {Physical Review E}\ }\textbf {\bibinfo {volume} {67}},\ \bibinfo {pages}
  {026126} (\bibinfo {year} {2003})}\BibitemShut {NoStop}%
\bibitem [{\citenamefont {Fr{\'e}chet}(1960)}]{frechet1960tableaux}%
  \BibitemOpen
  \bibfield  {author} {\bibinfo {author} {\bibfnamefont {Maurice}\ \bibnamefont
  {Fr{\'e}chet}},\ }\bibfield  {title} {\enquote {\bibinfo {title} {Sur les
  tableaux dont les marges et des bornes sont donn{\'e}es},}\ }\href@noop {}
  {\bibfield  {journal} {\bibinfo  {journal} {Revue de l'Institut international
  de statistique}\ ,\ \bibinfo {pages} {10--32}} (\bibinfo {year}
  {1960})}\BibitemShut {NoStop}%
\bibitem [{\citenamefont {Hoeffding}(1994)}]{hoeffding1994scale}%
  \BibitemOpen
  \bibfield  {author} {\bibinfo {author} {\bibfnamefont {Wassily}\ \bibnamefont
  {Hoeffding}},\ }\bibfield  {title} {\enquote {\bibinfo {title}
  {Scale—invariant correlation theory},}\ }in\ \href@noop {} {\emph {\bibinfo
  {booktitle} {The collected works of Wassily Hoeffding}}}\ (\bibinfo
  {publisher} {Springer},\ \bibinfo {year} {1994})\ pp.\ \bibinfo {pages}
  {57--107}\BibitemShut {NoStop}%
\bibitem [{\citenamefont {Jadidi}\ \emph {et~al.}(2017)\citenamefont {Jadidi},
  \citenamefont {Karimi}, \citenamefont {Lietz},\ and\ \citenamefont
  {Wagner}}]{jadidi2017gender}%
  \BibitemOpen
  \bibfield  {author} {\bibinfo {author} {\bibfnamefont {Mohsen}\ \bibnamefont
  {Jadidi}}, \bibinfo {author} {\bibfnamefont {Fariba}\ \bibnamefont {Karimi}},
  \bibinfo {author} {\bibfnamefont {Haiko}\ \bibnamefont {Lietz}}, \ and\
  \bibinfo {author} {\bibfnamefont {Claudia}\ \bibnamefont {Wagner}},\
  }\bibfield  {title} {\enquote {\bibinfo {title} {Gender disparities in
  science? dropout, productivity, collaborations and success of male and female
  computer scientists},}\ }\href@noop {} {\bibfield  {journal} {\bibinfo
  {journal} {Advances in Complex Systems}\ ,\ \bibinfo {pages} {1750011}}
  (\bibinfo {year} {2017})}\BibitemShut {NoStop}%
\bibitem [{\citenamefont {Lerman}\ \emph {et~al.}(2016)\citenamefont {Lerman},
  \citenamefont {Yan},\ and\ \citenamefont {Wu}}]{lerman2016majority}%
  \BibitemOpen
  \bibfield  {author} {\bibinfo {author} {\bibfnamefont {Kristina}\
  \bibnamefont {Lerman}}, \bibinfo {author} {\bibfnamefont {Xiaoran}\
  \bibnamefont {Yan}}, \ and\ \bibinfo {author} {\bibfnamefont {Xin-Zeng}\
  \bibnamefont {Wu}},\ }\bibfield  {title} {\enquote {\bibinfo {title} {The"
  majority illusion" in social networks},}\ }\href@noop {} {\bibfield
  {journal} {\bibinfo  {journal} {PloS one}\ }\textbf {\bibinfo {volume}
  {11}},\ \bibinfo {pages} {e0147617} (\bibinfo {year} {2016})}\BibitemShut
  {NoStop}%
\bibitem [{\citenamefont {Lee}\ \emph {et~al.}(2019)\citenamefont {Lee},
  \citenamefont {Karimi}, \citenamefont {Wagner}, \citenamefont {Jo},
  \citenamefont {Strohmaier},\ and\ \citenamefont
  {Galesic}}]{lee2017homophily}%
  \BibitemOpen
  \bibfield  {author} {\bibinfo {author} {\bibfnamefont {Eun}\ \bibnamefont
  {Lee}}, \bibinfo {author} {\bibfnamefont {Fariba}\ \bibnamefont {Karimi}},
  \bibinfo {author} {\bibfnamefont {Claudia}\ \bibnamefont {Wagner}}, \bibinfo
  {author} {\bibfnamefont {Hang-Hyun}\ \bibnamefont {Jo}}, \bibinfo {author}
  {\bibfnamefont {Markus}\ \bibnamefont {Strohmaier}}, \ and\ \bibinfo {author}
  {\bibfnamefont {Mirta}\ \bibnamefont {Galesic}},\ }\bibfield  {title}
  {\enquote {\bibinfo {title} {Homophily and minority-group size explain
  perception biases in social networks},}\ }\href@noop {} {\bibfield  {journal}
  {\bibinfo  {journal} {Nature Human Behaviour}\ } (\bibinfo {year}
  {2019})}\BibitemShut {NoStop}%
\bibitem [{\citenamefont {Fosdick}\ \emph {et~al.}(2018)\citenamefont
  {Fosdick}, \citenamefont {Larremore}, \citenamefont {Nishimura},\ and\
  \citenamefont {~}}]{fosdick2016configuring}%
  \BibitemOpen
  \bibfield  {author} {\bibinfo {author} {\bibfnamefont {Bailey~K}\
  \bibnamefont {Fosdick}}, \bibinfo {author} {\bibfnamefont {Daniel~B}\
  \bibnamefont {Larremore}}, \bibinfo {author} {\bibfnamefont {Joel}\
  \bibnamefont {Nishimura}}, \ and\ \bibinfo {author} {\bibfnamefont {Johan}\
  \bibnamefont {~}},\ }\bibfield  {title} {\enquote {\bibinfo {title}
  {Configuring random graph models with fixed degree sequences},}\ }\href@noop
  {} {\bibfield  {journal} {\bibinfo  {journal} {SIAM Review}\ }\textbf
  {\bibinfo {volume} {60}},\ \bibinfo {pages} {315--355} (\bibinfo {year}
  {2018})}\BibitemShut {NoStop}%
\bibitem [{\citenamefont {Park}\ and\ \citenamefont
  {Barab{\'a}si}(2007)}]{park2007distribution}%
  \BibitemOpen
  \bibfield  {author} {\bibinfo {author} {\bibfnamefont {Juyong}\ \bibnamefont
  {Park}}\ and\ \bibinfo {author} {\bibfnamefont {Albert-L{\'a}szl{\'o}}\
  \bibnamefont {Barab{\'a}si}},\ }\bibfield  {title} {\enquote {\bibinfo
  {title} {Distribution of node characteristics in complex networks},}\
  }\href@noop {} {\bibfield  {journal} {\bibinfo  {journal} {Proceedings of the
  National Academy of Sciences}\ }\textbf {\bibinfo {volume} {104}},\ \bibinfo
  {pages} {17916--17920} (\bibinfo {year} {2007})}\BibitemShut {NoStop}%
\bibitem [{\citenamefont {Yule}(1912)}]{yule1912on}%
  \BibitemOpen
  \bibfield  {author} {\bibinfo {author} {\bibfnamefont {G.~Udny}\ \bibnamefont
  {Yule}},\ }\bibfield  {title} {\enquote {\bibinfo {title} {On the methods of
  measuring association between two attributes},}\ }\href@noop {} {\bibfield
  {journal} {\bibinfo  {journal} {Journal of the Royal Statistical Society}\
  }\textbf {\bibinfo {volume} {75}},\ \bibinfo {pages} {579--652} (\bibinfo
  {year} {1912})}\BibitemShut {NoStop}%
\bibitem [{\citenamefont {Anscombe}(1973)}]{anscombe1973graphs}%
  \BibitemOpen
  \bibfield  {author} {\bibinfo {author} {\bibfnamefont {Francis~J}\
  \bibnamefont {Anscombe}},\ }\bibfield  {title} {\enquote {\bibinfo {title}
  {Graphs in statistical analysis},}\ }\href@noop {} {\bibfield  {journal}
  {\bibinfo  {journal} {The American Statistician}\ }\textbf {\bibinfo {volume}
  {27}},\ \bibinfo {pages} {17--21} (\bibinfo {year} {1973})}\BibitemShut
  {NoStop}%
\bibitem [{\citenamefont {Peel}\ \emph {et~al.}(2018)\citenamefont {Peel},
  \citenamefont {Delvenne},\ and\ \citenamefont
  {Lambiotte}}]{peel2018multiscale}%
  \BibitemOpen
  \bibfield  {author} {\bibinfo {author} {\bibfnamefont {Leto}\ \bibnamefont
  {Peel}}, \bibinfo {author} {\bibfnamefont {Jean-Charles}\ \bibnamefont
  {Delvenne}}, \ and\ \bibinfo {author} {\bibfnamefont {Renaud}\ \bibnamefont
  {Lambiotte}},\ }\bibfield  {title} {\enquote {\bibinfo {title} {Multiscale
  mixing patterns in networks},}\ }\href@noop {} {\bibfield  {journal}
  {\bibinfo  {journal} {Proceedings of the National Academy of Sciences}\
  }\textbf {\bibinfo {volume} {115}},\ \bibinfo {pages} {4057--4062} (\bibinfo
  {year} {2018})}\BibitemShut {NoStop}%
\bibitem [{\citenamefont {Cureton}(1959)}]{Cureton1959}%
  \BibitemOpen
  \bibfield  {author} {\bibinfo {author} {\bibfnamefont {Edward~E.}\
  \bibnamefont {Cureton}},\ }\bibfield  {title} {\enquote {\bibinfo {title}
  {Note on $\phi$/$\phi$max},}\ }\href {\doibase 10.1007/BF02289765} {\bibfield
   {journal} {\bibinfo  {journal} {Psychometrika}\ }\textbf {\bibinfo {volume}
  {24}},\ \bibinfo {pages} {89--91} (\bibinfo {year} {1959})}\BibitemShut
  {NoStop}%
\bibitem [{\citenamefont {Guilford}(1965)}]{guilford1965minimal}%
  \BibitemOpen
  \bibfield  {author} {\bibinfo {author} {\bibfnamefont {Joy~Paul}\
  \bibnamefont {Guilford}},\ }\bibfield  {title} {\enquote {\bibinfo {title}
  {The minimal phi coefficient and the maximal phi},}\ }\href@noop {}
  {\bibfield  {journal} {\bibinfo  {journal} {Educational and psychological
  measurement}\ }\textbf {\bibinfo {volume} {25}},\ \bibinfo {pages} {3--8}
  (\bibinfo {year} {1965})}\BibitemShut {NoStop}%
\bibitem [{\citenamefont {Cinelli}\ \emph {et~al.}(2017)\citenamefont
  {Cinelli}, \citenamefont {Ferraro},\ and\ \citenamefont
  {Iovanella}}]{cinelli2017structural}%
  \BibitemOpen
  \bibfield  {author} {\bibinfo {author} {\bibfnamefont {Matteo}\ \bibnamefont
  {Cinelli}}, \bibinfo {author} {\bibfnamefont {Giovanna}\ \bibnamefont
  {Ferraro}}, \ and\ \bibinfo {author} {\bibfnamefont {Antonio}\ \bibnamefont
  {Iovanella}},\ }\bibfield  {title} {\enquote {\bibinfo {title} {Structural
  bounds on the dyadic effect},}\ }\href {\doibase 10.1093/comnet/cnx002}
  {\bibfield  {journal} {\bibinfo  {journal} {Journal of Complex Networks}\
  }\textbf {\bibinfo {volume} {5}},\ \bibinfo {pages} {694--711} (\bibinfo
  {year} {2017})}\BibitemShut {NoStop}%
\bibitem [{\citenamefont {Cinelli}\ \emph {et~al.}(2019)\citenamefont
  {Cinelli}, \citenamefont {Ferraro},\ and\ \citenamefont
  {Iovanella}}]{cinelli2019evaluating}%
  \BibitemOpen
  \bibfield  {author} {\bibinfo {author} {\bibfnamefont {Matteo}\ \bibnamefont
  {Cinelli}}, \bibinfo {author} {\bibfnamefont {Giovanna}\ \bibnamefont
  {Ferraro}}, \ and\ \bibinfo {author} {\bibfnamefont {Antonio}\ \bibnamefont
  {Iovanella}},\ }\bibfield  {title} {\enquote {\bibinfo {title} {Evaluating
  relevance and redundancy to quantify how binary node metadata interplay with
  the network structure},}\ }\href@noop {} {\bibfield  {journal} {\bibinfo
  {journal} {Scientific reports}\ }\textbf {\bibinfo {volume} {9}} (\bibinfo
  {year} {2019})}\BibitemShut {NoStop}%
\bibitem [{\citenamefont {Di~Paola}\ \emph {et~al.}(2012)\citenamefont
  {Di~Paola}, \citenamefont {De~Ruvo}, \citenamefont {Paci}, \citenamefont
  {Santoni},\ and\ \citenamefont {Giuliani}}]{di2012protein}%
  \BibitemOpen
  \bibfield  {author} {\bibinfo {author} {\bibfnamefont {Luisa}\ \bibnamefont
  {Di~Paola}}, \bibinfo {author} {\bibfnamefont {Micol}\ \bibnamefont
  {De~Ruvo}}, \bibinfo {author} {\bibfnamefont {Paola}\ \bibnamefont {Paci}},
  \bibinfo {author} {\bibfnamefont {Daniele}\ \bibnamefont {Santoni}}, \ and\
  \bibinfo {author} {\bibfnamefont {Alessandro}\ \bibnamefont {Giuliani}},\
  }\bibfield  {title} {\enquote {\bibinfo {title} {Protein contact networks: an
  emerging paradigm in chemistry},}\ }\href@noop {} {\bibfield  {journal}
  {\bibinfo  {journal} {Chemical Reviews}\ }\textbf {\bibinfo {volume} {113}},\
  \bibinfo {pages} {1598--1613} (\bibinfo {year} {2012})}\BibitemShut {NoStop}%
\bibitem [{\citenamefont {Lusseau}\ and\ \citenamefont
  {Newman}(2004)}]{lusseau2004identifying}%
  \BibitemOpen
  \bibfield  {author} {\bibinfo {author} {\bibfnamefont {David}\ \bibnamefont
  {Lusseau}}\ and\ \bibinfo {author} {\bibfnamefont {Mark~EJ}\ \bibnamefont
  {Newman}},\ }\bibfield  {title} {\enquote {\bibinfo {title} {Identifying the
  role that animals play in their social networks},}\ }\href@noop {} {\bibfield
   {journal} {\bibinfo  {journal} {Proceedings of the Royal Society of London
  B: Biological Sciences}\ }\textbf {\bibinfo {volume} {271}},\ \bibinfo
  {pages} {S477--S481} (\bibinfo {year} {2004})}\BibitemShut {NoStop}%
\bibitem [{\citenamefont {McPherson}\ \emph {et~al.}(2001)\citenamefont
  {McPherson}, \citenamefont {Smith-Lovin},\ and\ \citenamefont
  {Cook}}]{mcpherson2001birds}%
  \BibitemOpen
  \bibfield  {author} {\bibinfo {author} {\bibfnamefont {Miller}\ \bibnamefont
  {McPherson}}, \bibinfo {author} {\bibfnamefont {Lynn}\ \bibnamefont
  {Smith-Lovin}}, \ and\ \bibinfo {author} {\bibfnamefont {James~M}\
  \bibnamefont {Cook}},\ }\bibfield  {title} {\enquote {\bibinfo {title} {Birds
  of a feather: Homophily in social networks},}\ }\href@noop {} {\bibfield
  {journal} {\bibinfo  {journal} {Annual review of sociology}\ }\textbf
  {\bibinfo {volume} {27}},\ \bibinfo {pages} {415--444} (\bibinfo {year}
  {2001})}\BibitemShut {NoStop}%
\bibitem [{\citenamefont {Laniado}\ \emph {et~al.}(2016)\citenamefont
  {Laniado}, \citenamefont {Volkovich}, \citenamefont {Kappler},\ and\
  \citenamefont {Kaltenbrunner}}]{laniado2016gender}%
  \BibitemOpen
  \bibfield  {author} {\bibinfo {author} {\bibfnamefont {David}\ \bibnamefont
  {Laniado}}, \bibinfo {author} {\bibfnamefont {Yana}\ \bibnamefont
  {Volkovich}}, \bibinfo {author} {\bibfnamefont {Karolin}\ \bibnamefont
  {Kappler}}, \ and\ \bibinfo {author} {\bibfnamefont {Andreas}\ \bibnamefont
  {Kaltenbrunner}},\ }\bibfield  {title} {\enquote {\bibinfo {title} {Gender
  homophily in online dyadic and triadic relationships},}\ }\href {\doibase
  10.1140/epjds/s13688-016-0080-6} {\bibfield  {journal} {\bibinfo  {journal}
  {EPJ Data Science}\ }\textbf {\bibinfo {volume} {5}},\ \bibinfo {pages} {19}
  (\bibinfo {year} {2016})}\BibitemShut {NoStop}%
\bibitem [{\citenamefont {Centola}(2011)}]{centola2011anexperimental}%
  \BibitemOpen
  \bibfield  {author} {\bibinfo {author} {\bibfnamefont {Damon}\ \bibnamefont
  {Centola}},\ }\bibfield  {title} {\enquote {\bibinfo {title} {An experimental
  study of homophily in the adoption of health behavior},}\ }\href {\doibase
  10.1126/science.1207055} {\bibfield  {journal} {\bibinfo  {journal}
  {Science}\ }\textbf {\bibinfo {volume} {334}},\ \bibinfo {pages} {1269--1272}
  (\bibinfo {year} {2011})},\ \Eprint
  {http://arxiv.org/abs/http://science.sciencemag.org/content/334/6060/1269.full.pdf}
  {http://science.sciencemag.org/content/334/6060/1269.full.pdf} \BibitemShut
  {NoStop}%
\bibitem [{\citenamefont {Cunningham}\ \emph {et~al.}(2012)\citenamefont
  {Cunningham}, \citenamefont {Vaquera}, \citenamefont {Maturo},\ and\
  \citenamefont {Narayan}}]{cunningham2012isthereevidence}%
  \BibitemOpen
  \bibfield  {author} {\bibinfo {author} {\bibfnamefont {Solveig~A.}\
  \bibnamefont {Cunningham}}, \bibinfo {author} {\bibfnamefont {Elizabeth}\
  \bibnamefont {Vaquera}}, \bibinfo {author} {\bibfnamefont {Claire~C.}\
  \bibnamefont {Maturo}}, \ and\ \bibinfo {author} {\bibfnamefont
  {K.M.~Venkat}\ \bibnamefont {Narayan}},\ }\bibfield  {title} {\enquote
  {\bibinfo {title} {Is there evidence that friends influence body weight? a
  systematic review of empirical research},}\ }\href@noop {} {\bibfield
  {journal} {\bibinfo  {journal} {Social Science \& Medicine}\ }\textbf
  {\bibinfo {volume} {75}},\ \bibinfo {pages} {1175--1183} (\bibinfo {year}
  {2012})}\BibitemShut {NoStop}%
\bibitem [{\citenamefont {Rosenquist}\ \emph {et~al.}(2010)\citenamefont
  {Rosenquist}, \citenamefont {Murabito}, \citenamefont {Fowler},\ and\
  \citenamefont {Christakis}}]{rosenquist2010spread}%
  \BibitemOpen
  \bibfield  {author} {\bibinfo {author} {\bibfnamefont {J~Niels}\ \bibnamefont
  {Rosenquist}}, \bibinfo {author} {\bibfnamefont {Joanne}\ \bibnamefont
  {Murabito}}, \bibinfo {author} {\bibfnamefont {James~H}\ \bibnamefont
  {Fowler}}, \ and\ \bibinfo {author} {\bibfnamefont {Nicholas~A}\ \bibnamefont
  {Christakis}},\ }\bibfield  {title} {\enquote {\bibinfo {title} {The spread
  of alcohol consumption behavior in a large social network},}\ }\href@noop {}
  {\bibfield  {journal} {\bibinfo  {journal} {Annals of internal medicine}\
  }\textbf {\bibinfo {volume} {152}},\ \bibinfo {pages} {426--433} (\bibinfo
  {year} {2010})}\BibitemShut {NoStop}%
\bibitem [{\citenamefont {Altenburger}\ and\ \citenamefont
  {Ugander}(2018)}]{altenburger2017bias}%
  \BibitemOpen
  \bibfield  {author} {\bibinfo {author} {\bibfnamefont {Kristen~M}\
  \bibnamefont {Altenburger}}\ and\ \bibinfo {author} {\bibfnamefont {Johan}\
  \bibnamefont {Ugander}},\ }\bibfield  {title} {\enquote {\bibinfo {title}
  {Monophily in social networks introduces similarity among
  friends-of-friends},}\ }\href@noop {} {\bibfield  {journal} {\bibinfo
  {journal} {Nature human behaviour}\ }\textbf {\bibinfo {volume} {2}},\
  \bibinfo {pages} {284} (\bibinfo {year} {2018})}\BibitemShut {NoStop}%
\bibitem [{\citenamefont {Traud}\ \emph {et~al.}(2012)\citenamefont {Traud},
  \citenamefont {Mucha},\ and\ \citenamefont {Porter}}]{traud2012social}%
  \BibitemOpen
  \bibfield  {author} {\bibinfo {author} {\bibfnamefont {Amanda~L}\
  \bibnamefont {Traud}}, \bibinfo {author} {\bibfnamefont {Peter~J}\
  \bibnamefont {Mucha}}, \ and\ \bibinfo {author} {\bibfnamefont {Mason~A}\
  \bibnamefont {Porter}},\ }\bibfield  {title} {\enquote {\bibinfo {title}
  {Social structure of facebook networks},}\ }\href@noop {} {\bibfield
  {journal} {\bibinfo  {journal} {Physica A: Statistical Mechanics and its
  Applications}\ }\textbf {\bibinfo {volume} {391}},\ \bibinfo {pages}
  {4165--4180} (\bibinfo {year} {2012})}\BibitemShut {NoStop}%
\bibitem [{\citenamefont {Peel}\ \emph {et~al.}(2017)\citenamefont {Peel},
  \citenamefont {Larremore},\ and\ \citenamefont {Clauset}}]{peel2017ground}%
  \BibitemOpen
  \bibfield  {author} {\bibinfo {author} {\bibfnamefont {Leto}\ \bibnamefont
  {Peel}}, \bibinfo {author} {\bibfnamefont {Daniel~B}\ \bibnamefont
  {Larremore}}, \ and\ \bibinfo {author} {\bibfnamefont {Aaron}\ \bibnamefont
  {Clauset}},\ }\bibfield  {title} {\enquote {\bibinfo {title} {The ground
  truth about metadata and community detection in networks},}\ }\href@noop {}
  {\bibfield  {journal} {\bibinfo  {journal} {Science advances}\ }\textbf
  {\bibinfo {volume} {3}},\ \bibinfo {pages} {e1602548} (\bibinfo {year}
  {2017})}\BibitemShut {NoStop}%
\bibitem [{\citenamefont {van Hooff}\ and\ \citenamefont
  {Wensing}(1987)}]{van1987dominance}%
  \BibitemOpen
  \bibfield  {author} {\bibinfo {author} {\bibfnamefont {Jan~A.R.A.M.}\
  \bibnamefont {van Hooff}}\ and\ \bibinfo {author} {\bibfnamefont {Joep~A.B.}\
  \bibnamefont {Wensing}},\ }\bibfield  {title} {\enquote {\bibinfo {title}
  {Dominance and its behavioral measures in a captive wolf pack.}}\ }in\
  \href@noop {} {\emph {\bibinfo {booktitle} {Man and Wolf: Advances, Issues,
  and Problems in Captive Wolf Research}}},\ \bibinfo {editor} {edited by\
  \bibinfo {editor} {\bibfnamefont {H.}~\bibnamefont {Frank}}}\ (\bibinfo
  {publisher} {Dr W Junk Publishers},\ \bibinfo {year} {1987})\ pp.\ \bibinfo
  {pages} {219--252}\BibitemShut {NoStop}%
\bibitem [{\citenamefont {Cohen}(1960)}]{cohen1960coefficient}%
  \BibitemOpen
  \bibfield  {author} {\bibinfo {author} {\bibfnamefont {Jacob}\ \bibnamefont
  {Cohen}},\ }\bibfield  {title} {\enquote {\bibinfo {title} {A coefficient of
  agreement for nominal scales},}\ }\href@noop {} {\bibfield  {journal}
  {\bibinfo  {journal} {Educational and psychological measurement}\ }\textbf
  {\bibinfo {volume} {20}},\ \bibinfo {pages} {37--46} (\bibinfo {year}
  {1960})}\BibitemShut {NoStop}%
\bibitem [{\citenamefont {Ugander}\ \emph {et~al.}(2011)\citenamefont
  {Ugander}, \citenamefont {Karrer}, \citenamefont {Backstrom},\ and\
  \citenamefont {Marlow}}]{ugander2011anatomy}%
  \BibitemOpen
  \bibfield  {author} {\bibinfo {author} {\bibfnamefont {Johan}\ \bibnamefont
  {Ugander}}, \bibinfo {author} {\bibfnamefont {Brian}\ \bibnamefont {Karrer}},
  \bibinfo {author} {\bibfnamefont {Lars}\ \bibnamefont {Backstrom}}, \ and\
  \bibinfo {author} {\bibfnamefont {Cameron}\ \bibnamefont {Marlow}},\
  }\bibfield  {title} {\enquote {\bibinfo {title} {The anatomy of the facebook
  social graph},}\ }\href@noop {} {\bibfield  {journal} {\bibinfo  {journal}
  {arXiv preprint arXiv:1111.4503}\ } (\bibinfo {year} {2011})}\BibitemShut
  {NoStop}%
\bibitem [{\citenamefont {Jacobs}\ \emph {et~al.}(2015)\citenamefont {Jacobs},
  \citenamefont {Way}, \citenamefont {Ugander},\ and\ \citenamefont
  {Clauset}}]{jacobs2015assembling}%
  \BibitemOpen
  \bibfield  {author} {\bibinfo {author} {\bibfnamefont {Abigail~Z}\
  \bibnamefont {Jacobs}}, \bibinfo {author} {\bibfnamefont {Samuel~F}\
  \bibnamefont {Way}}, \bibinfo {author} {\bibfnamefont {Johan}\ \bibnamefont
  {Ugander}}, \ and\ \bibinfo {author} {\bibfnamefont {Aaron}\ \bibnamefont
  {Clauset}},\ }\bibfield  {title} {\enquote {\bibinfo {title} {Assembling
  thefacebook: Using heterogeneity to understand online social network
  assembly},}\ }in\ \href@noop {} {\emph {\bibinfo {booktitle} {Proceedings of
  the ACM Web Science Conference}}}\ (\bibinfo {organization} {ACM},\ \bibinfo
  {year} {2015})\ p.~\bibinfo {pages} {18}\BibitemShut {NoStop}%
\bibitem [{\citenamefont {Dorogovtsev}\ \emph {et~al.}(2010)\citenamefont
  {Dorogovtsev}, \citenamefont {Ferreira}, \citenamefont {Goltsev},\ and\
  \citenamefont {Mendes}}]{dorogovtsev2010zero}%
  \BibitemOpen
  \bibfield  {author} {\bibinfo {author} {\bibfnamefont {SN}~\bibnamefont
  {Dorogovtsev}}, \bibinfo {author} {\bibfnamefont {AL}~\bibnamefont
  {Ferreira}}, \bibinfo {author} {\bibfnamefont {AV}~\bibnamefont {Goltsev}}, \
  and\ \bibinfo {author} {\bibfnamefont {JFF}\ \bibnamefont {Mendes}},\
  }\bibfield  {title} {\enquote {\bibinfo {title} {Zero pearson coefficient for
  strongly correlated growing trees},}\ }\href@noop {} {\bibfield  {journal}
  {\bibinfo  {journal} {Physical Review E}\ }\textbf {\bibinfo {volume} {81}},\
  \bibinfo {pages} {031135} (\bibinfo {year} {2010})}\BibitemShut {NoStop}%
\bibitem [{\citenamefont {Newman}(2010)}]{newman2010networks}%
  \BibitemOpen
  \bibfield  {author} {\bibinfo {author} {\bibfnamefont {Mark~EJ}\ \bibnamefont
  {Newman}},\ }\href@noop {} {\emph {\bibinfo {title} {Networks: An
  Introduction}}}\ (\bibinfo  {publisher} {Oxford University Press},\ \bibinfo
  {address} {New York},\ \bibinfo {year} {2010})\BibitemShut {NoStop}%
\bibitem [{\citenamefont {Bianconi}\ \emph {et~al.}(2009)\citenamefont
  {Bianconi}, \citenamefont {Pin},\ and\ \citenamefont
  {Marsili}}]{bianconi2009assessing}%
  \BibitemOpen
  \bibfield  {author} {\bibinfo {author} {\bibfnamefont {Ginestra}\
  \bibnamefont {Bianconi}}, \bibinfo {author} {\bibfnamefont {Paolo}\
  \bibnamefont {Pin}}, \ and\ \bibinfo {author} {\bibfnamefont {Matteo}\
  \bibnamefont {Marsili}},\ }\bibfield  {title} {\enquote {\bibinfo {title}
  {Assessing the relevance of node features for network structure},}\
  }\href@noop {} {\bibfield  {journal} {\bibinfo  {journal} {Proceedings of the
  National Academy of Sciences}\ }\textbf {\bibinfo {volume} {106}},\ \bibinfo
  {pages} {11433--11438} (\bibinfo {year} {2009})}\BibitemShut {NoStop}%
\bibitem [{\citenamefont {Yang}\ \emph {et~al.}(2017)\citenamefont {Yang},
  \citenamefont {Pan},\ and\ \citenamefont {Zhou}}]{yang2017lower}%
  \BibitemOpen
  \bibfield  {author} {\bibinfo {author} {\bibfnamefont {Dan}\ \bibnamefont
  {Yang}}, \bibinfo {author} {\bibfnamefont {Liming}\ \bibnamefont {Pan}}, \
  and\ \bibinfo {author} {\bibfnamefont {Tao}\ \bibnamefont {Zhou}},\
  }\bibfield  {title} {\enquote {\bibinfo {title} {Lower bound of assortativity
  coefficient in scale-free networks},}\ }\href {\doibase 10.1063/1.4976030}
  {\bibfield  {journal} {\bibinfo  {journal} {Chaos: An Interdisciplinary
  Journal of Nonlinear Science}\ }\textbf {\bibinfo {volume} {27}},\ \bibinfo
  {pages} {033113} (\bibinfo {year} {2017})},\ \Eprint
  {http://arxiv.org/abs/https://doi.org/10.1063/1.4976030}
  {https://doi.org/10.1063/1.4976030} \BibitemShut {NoStop}%
\bibitem [{\citenamefont {Freeman}(1978)}]{freeman1978segregation}%
  \BibitemOpen
  \bibfield  {author} {\bibinfo {author} {\bibfnamefont {Linton~C}\
  \bibnamefont {Freeman}},\ }\bibfield  {title} {\enquote {\bibinfo {title}
  {Segregation in social networks},}\ }\href@noop {} {\bibfield  {journal}
  {\bibinfo  {journal} {Sociological Methods \& Research}\ }\textbf {\bibinfo
  {volume} {6}},\ \bibinfo {pages} {411--429} (\bibinfo {year}
  {1978})}\BibitemShut {NoStop}%
\bibitem [{\citenamefont {Estrada}(2011)}]{estrada2011combinatorial}%
  \BibitemOpen
  \bibfield  {author} {\bibinfo {author} {\bibfnamefont {Ernesto}\ \bibnamefont
  {Estrada}},\ }\bibfield  {title} {\enquote {\bibinfo {title} {Combinatorial
  study of degree assortativity in networks},}\ }\href {\doibase
  10.1103/PhysRevE.84.047101} {\bibfield  {journal} {\bibinfo  {journal} {Phys.
  Rev. E}\ }\textbf {\bibinfo {volume} {84}},\ \bibinfo {pages} {047101}
  (\bibinfo {year} {2011})}\BibitemShut {NoStop}%
\bibitem [{\citenamefont {Schenkel}(1967)}]{schenkel1967submission}%
  \BibitemOpen
  \bibfield  {author} {\bibinfo {author} {\bibfnamefont {Rudolf}\ \bibnamefont
  {Schenkel}},\ }\bibfield  {title} {\enquote {\bibinfo {title} {Submission:
  its features and function in the wolf and dog},}\ }\href@noop {} {\bibfield
  {journal} {\bibinfo  {journal} {American Zoologist}\ }\textbf {\bibinfo
  {volume} {7}},\ \bibinfo {pages} {319--329} (\bibinfo {year}
  {1967})}\BibitemShut {NoStop}%
\end{thebibliography}

%

\appendix

\section{Methods}
\label{Appendix}

\subsection{Bounds for the $\phi$-coefficient}
\label{sec_phi_bounds}
The bounds for the $\phi$-coefficient depend on the marginals $a_0, a_1, b_0, b_1$ of the contingency table in Eq.~\eqref{eq_contingency}~\cite{guilford1965minimal}.
We start by deriving an alternative expression for the numerator of the $\phi$-coefficient.
\begin{align}
  e_{11} - a_1b_1 
  & = e_{11} - (e_{11} + e_{01})(e_{11} + e_{10}) \notag \\
  & = e_{11} - (e_{11}^2 + e_{11}e_{01} + e_{11}e_{10} +  e_{01}e_{10}) \notag \\
  & = e_{11}(1 - e_{11} - e_{01} - e_{10}) -  e_{01}e_{10}\notag \\
  & = e_{11}e_{00} -  e_{01}e_{10} \enspace , \notag 
\end{align}
which gives us the alternative form for $\phi$,
\begin{equation}
 \phi = \frac{e_{11}e_{00} -  e_{01}e_{10}}{\sqrt{a_{1}a_{0}b_{1}a_{0}}} \enspace .
 \label{eq_alt_phi}
\end{equation}

From Eq.~\eqref{eq_alt_phi} we can easily see that the $\phi$-coefficient is at its minimum when either $e_{00} = 0$ and/or $e_{11}=0$. Here we will assume that $e_{00} \leq e_{11}$, so setting $e_{00} = 0$ means that $e_{01} = a_0$ and $e_{10} = b_0$. Then,
\begin{equation}
  \phi_{\min} = - \frac{e_{01}e_{10}}{\sqrt{a_{1}a_{0}b_{1}a_{0}}} = - \sqrt{\frac{a_0b_0}{a_1b_1}} \enspace .
\end{equation}

Similarly for the maximum of $\phi$, either $e_{01} = 0$ and/or $e_{01}=0$. So, if $e_{01}=0$ then:
\begin{equation}
  \phi_{\max} = \sqrt{\frac{a_0b_1}{a_1b_0}} \enspace .
\end{equation}

\subsection{Bounding the edge counts in the metadata-graph space}
\label{EC}
Here we summarize the bounds introduced in \cite{cinelli2017structural}. Given a degree sequence $D_G$, by using the quantities $n_1$ and $n_0$ which identify the amount of nodes with features 1 and 0 respectively, it is possible to define its head $D_G^H(n_1)$ or $D_G^H(n_0)$ and its tail $D_G^T(n_1)$ or $D_G^T(n_0)$ such that $D_G = D_G^H(n_1) \cup D_G^T(n_0)$ or  $D_G = D_G^H(n_0) \cup D_G^T(n_1)$.

Considering these partitions, the first upper bound $m_{11}^u$, is based on the fact that, especially in sparse networks, large cliques may be rare substructures. 
Therefore, using $D_G$ we check whether $G$ can actually contain a complete subgraph of size $n_1$ (i.e. if $D_G^H(n_1)$ satisfies the necessary condition for the realization of a clique). If not, we take into account the densest hypothetical substructure that could be realized using the degree sequence of $G$. In Eq.~\eqref{ubm11}, the first term is the number of links in the network, the second term is the number of links in a clique of size $n_1$, while  the third term is the number of links in the sub-graph with $n_1$ nodes and maximum degree-sum (i.e., with degree sequence $D_G^H(n_1)$).
\begin{equation}\label{ubm11}
m_{11}^u = \min\Bigg( m, \binom{n_1} {2}, \bigg\lceil \sum_{i \in D_G^H(n_1)}{\frac {\min(d_{i}, n_1 - 1)}{2}} \bigg\rceil \Bigg)     \enspace ,
\end{equation}
In the second upper bound, $m_{10}^u$, we check if $G$ can contain a complete bipartite subgraph with partitions size $n_1$ and $n_0$. If not, we consider a set of stars made of the first $n_1$ elements of $D_G$ if $n_1 < n_0$ or made of the first $n_0$ elements of $D_G$ if $n_0 < n_1$. In Eq.~\eqref{ubm10} the first term is the number of links in the network, the second term is the number of links in a bipartite graph with partitions of size $n_1$ and $n_0$, while  the third term is the minimum between the number of $m_{10}$ deriving from the degree partition $D_G^H(n_1) \cup D_G^T(n_0)$ and the number of $m_{10}$
deriving from the degree partition $D_G^H(n_0) \cup D_G^T(n_1)$.
\begin{eqnarray}\label{ubm10}
m_{10}^u = \min\Bigg( m, n_1 n_0, \min\bigg( \sum_{i \in D_G^H(n_1)}{\min(d_{i}, n_0)} , \sum_{i \in D_G^H(n_0)}{\min(d_{i}, n_1)}  \bigg)\Bigg) \, .
\end{eqnarray}
The first lower bound, $m_{11}^l$, considers the partition $D_{G}^{T}(n_1)$ and the minimum residual degree of its elements (when $>0$), which is exploited in order to realize the minimum $m_{11}$. Since most real networks are sparse, this bound is effective mainly in the case of unbalanced partitions and of dense networks. The second term of Eq.~\eqref{lbm11} counts the minimum number of links among the $n_1$ nodes in the graph deriving from the partition $D_G^H(n_0) \cup D_G^T(n_1)$, i.e. the amount of $m_{11}$ which is realizable from the residual degree of the partition $D_G^T(n_1)$.
\begin{equation}\label{lbm11}
m_{11}^l = \max \Bigg(0, \bigg \lfloor \frac{\sum_{i \in D_G^T(n_1)} d_i -  \sum_{i \in D_G^H(n_0)} d_i }{2} \bigg \rfloor\Bigg)
\end{equation}
The second lower bound, $m_{10}^l$, considers that the lower $m_{10}$ occurs in the case of a bisected network (i.e. a network with two separated components). Thus, if the degree sum in  $D_{G}^{T}(n_1)$ overcomes the degree sum in a clique of size $n_1$ then we guarantee the presence of some $m_{10}$. Considering that any connected realization with $n_1 \neq \{0, n\}$ has at least one $m_{10}$, the second term of Eq.~\ref{lbm10} counts the minimum number of links between the $n_1$ and $n_0$ in the case the $n_1$ are arranged into a clique.
\begin{equation}\label{lbm10}
\begin{array}{rl}
m_{10}^l = &
  \begin{cases}
    0                                                                                & \! \text{if } n_1 = 0, n\\ 
    \max \bigg(1; \sum_{i \in D_G^T(n_1)} d_i - n_1(n_1 -1)\bigg)  &\!  \text{if } 0 < n_1 < n  \\
  \end{cases}
\end{array}
\end{equation}
%
%
The bounds to $m_{00}$ can be obtained using the same rationale as that of $m_{11}$.
\subsubsection{Improvements to lower bounds in the metadata-graph space}
\label{ILB}
The lower bound to the intra-partition links is $m_{11}^l$. It can be initially improved by correcting the term $\sum_{i \in D_G^H(n_0)} d_i $. This term keeps the bound low especially in the case of unbalanced partitions and in the case of heavy tailed and sparse networks (i.e. when the degree sum of $D_G^H(n_0)$ has a high value because of the presence of hubs). Knowing the size of the two partitions, the second term in $m_{11}^l$ can  be written as: $\sum_{i \in D_G^H(n_0)} \min(d_i, n_1)$.

Indeed, any node in $n_0$, despite its degree, can be connected at most to other $n_1$ nodes in a different partition. Consequently the residual degree of the nodes in $D_G^T(n_1)$ can be exploited for the realization of $m_{11}$.

Therefore:
\begin{equation}
 m_{11}^l = \max \Bigg(0, \bigg \lfloor \frac{\sum_{i \in D_G^T(n_1)} d_i -  \sum_{i \in D_G^H(n_0)} \min(d_i, n_1) }{2} \bigg \rfloor\Bigg) \, .
\end{equation}
The bound to the inter-partition links is $m_{10}^l$. The first extension consists in making the bound symmetrical by adding the term $\sum_{i \in D_G^T(n_0)}d_i - n_0(n_0 -1)$ and in noticing that such term can be written in a more efficient way as $\sum_{i \in D_G^T(n_0)} \max (0, d_i - (n_0 -1))$. As shown in \cite{cinelli2017structural}, the current bound works better in the case of dense networks since, when $n_1$ becomes larger, the nodes in $D_G^T(n_1)$ may still have a residual degree which is higher than the degree of the nodes in a clique of size $n_1$ (i.e. certain elements in $D_G^T(n_1)$ have degree greater than $n_1 -1$). Conversely, if the considered network is relatively sparse we may not be able to provide a lower bound to $m_{10}$ which is greater than zero even for very low values of $n_1$.

Therefore, given that $n_1 + n_0 = n$, when $n_1$ increases we should also try to bound $m_{10}$ by supposing a realization in the tail of $D_G$ that involves $n_0$ nodes. Thus, the symmetrical version of $m_{10}^l$ comprises the term $\sum_{i \in D_G^T(n_0)}d_i - n_0(n_0 -1)$.

An additional improvement, possibly more appropriate in the case of heavy tailed and sparse degree sequences, derives from the following consideration: called $D_G(n_1)$ and $D_G(n_0)$ two arbitrary partitions of $D_G$, any element in $D_G(n_1)$ ($D_G(n_0))$ can be connected at most to other $n_1 -1$ ($n_0 -1$) ones in the same partition. Thus, any element in $D_G(n_1)$ ($D_G(n_0)$) can be involved in at least $d_i - (n_1 -1)$ ($d_i - (n_0-1)$) intra-partition links.

Given a certain arbitrary partition of $D_G = D_G(n_1) \cup D_G(n_0)$, the minimum amount of $m_{10}$ that can be realized is:
$m_{10} = \frac{1}{2} \bigg(\sum_{i \in D_G(n_1)} d_i - (n_1 -1) + \sum_{j \in D_G(n_0)} d_j - (n_0 -1) \bigg)$.

In the case $n_1 > n_0$ the following relation holds: $\frac{1}{2} \bigg(\sum_{i \in D_G(n_1)} d_i - (n_1 -1) + \sum_{j \in D_G(n_0)} d_j - (n_0 -1) \bigg) \geq \frac{1}{2} \bigg(\sum_{i \in D_G(n_1)} d_i - (n_1 -1) + \sum_{j \in D_G(n_0)} d_j - (n_1 -1) \bigg)$.
   
The second term of such a relation assumes, in order to provide a lower bound to $m_{10}$, that any element of $D_G$ has the lowest possible residual degree for the realization of $m_{10}$. Obviously, the quantity $d_i - (n_1 -1)$ has to be greater than 0 for each $i$ and the second term of the previous inequality represents the lowest possible sum of residual degrees of any arbitrary partition, in the case $n_1 > n_0$. 
Thus, the previous relation can be written as: $\frac{1}{2} \sum_{i=1}^n \max(0, d_i - (n_1 -1))$. Finally, $m_{10}^l$ can be expressed as:
\begin{itemize}
\item if $n_1 = 0, n$
\begin{equation}
m_{10}^l = 0
\end{equation}
\item if $n_1 > n_0$
\begin{equation}
m_{10}^l = \max \bigg(1; \sum_{i \in D_G^T(n_1)} \max (0, d_i - (n_1 -1)); \sum_{i \in D_G^T(n_0)} \max (0, d_i - (n_0 -1));
                        \big \lfloor \frac{1}{2} \sum_{i=1}^n \max(0, d_i - (n_1 -1)) \big \rfloor \bigg)
\end{equation}
\item if $n_1 \leq n_0$
\begin{equation}
m_{10}^l =  \max \bigg(1; \sum_{i \in D_G^T(n_1)} \max (0, d_i - (n_1 -1)); \sum_{i \in D_G^T(n_0)} \max (0, d_i - (n_0 -1));
                          \big \lfloor \frac{1}{2} \sum_{i=1}^n \max(0, d_i - (n_0 -1)) \big \rfloor \bigg)
\end{equation}
\end{itemize}

Or in a more compact way when $n_1 \neq 0, n$:
\begin{equation}
m_{10}^l = \max \bigg(1; \sum_{i \in D_G^T(n_1)} \max (0, d_i - (n_1 -1)); \sum_{i \in D_G^T(n_0)} \max (0, d_i - (n_0 -1));
                          \big \lfloor \frac{1}{2} \sum_{i=1}^n \max(0, d_i - (\max (n_1, n_0) -1)) \big \rfloor \bigg)  
\end{equation}

\subsection{Bounding the edge counts in the graph space}
\label{ECGS}

We consider the graph space into which the degree sequence $D_G$ and the vector of binary node metadata are both fixed. In such a case, we say that $D_G=D_G(n_1) \cup D_G(n_0)$ which represents the current partition of the considered degree sequence, given the node metadata assignment. Therefore, we can exploit the combinatorial bounds of the metadata-graph space in order to bound the different edge counts in the graph space. The rationale behind the bounds remain the same as well as the formulas (presented in Sections \ref{EC} and \ref{ILB}) which can be, however, contracted as we can't leverage, within the graph space, the different ways of partitioning $D_G$. Therefore, the bounds in the graph space can be written as:
\begin{equation}\label{ubm11c}
m_{11}^u = \min\Bigg( m, \binom{n_1} {2}, \bigg\lceil \sum_{i \in D_G(n_1)}{\frac {\min(d_{i}, n_1 - 1)}{2}} \bigg\rceil \Bigg) \, ,    
\end{equation}
\begin{eqnarray}\label{ubm10c}
m_{10}^u = \min\Bigg( m, n_1 n_0, \min\bigg( \sum_{i \in D_G(n_1)}{\min(d_{i}, n_0)} , \sum_{i \in D_G(n_0)}{\min(d_{i}, n_1)}  \bigg)\Bigg) \, ,  
\end{eqnarray}
\begin{equation}
 m_{11}^l = \max \Bigg(0, \bigg \lfloor \frac{\sum_{i \in D_G(n_1)} d_i -  \sum_{i \in D_G(n_0)} \min(d_i, n_1) }{2} \bigg \rfloor\Bigg) \, ,
\end{equation}
\begin{equation}
m_{10}^l = \max \bigg(1; \sum_{i \in D_G(n_1)} \max (0, d_i - (n_1 -1)); \sum_{i \in D_G(n_0)} \max (0, d_i - (n_0 -1)) \bigg) \, .
\end{equation}

\subsection{Swap of node metadata}
\label{switch}

In order to approximate the maximum and minimum values of binary assortativity in the metadata space we use the following heuristic procedure which provides admissible solutions to the graph bisection problem also in the case of unbalanced partitions.

\begin{enumerate}

\item Take into account the network, the metadata vector $c$ and compute $r^{current}$

\item Take into account two randomly chosen entries of $c$, called $c_i$ and $c_j$, such that $c_i = 1$ and $c_j = 0$ (or viceversa)  

\item Swap the values of $c_i$ and $c_j$ and compute $r^{swap}$

\begin{itemize}

\item In the case of assortativity maximization:

if $r^{swap}>r^{current}$ then the switch is accepted and $r^{current}=r^{swap}$

if $r^{swap} \leq r^{current}$ then with probability $p=0.001$ 
the swap is accepted, and $r^{current}=r^{swap}$

\item In the case of assortativity minimization:

if $r^{swap}<r^{current}$ then the swap is accepted and $r^{current}=r^{swap}$

if $r^{swap} \geq r^{current}$ then with probability $p$ the swap is accepted, and $r^{current}=r^{swap}$
\end{itemize}

\end{enumerate}

Steps 2 and 3 of procedure are iterated several times and different repetitions are performed.

\subsection{Freeman's Segregation}
\label{segregation}

By using the notation of \cite{freeman1978segregation}, segregation $S$ can be expressed starting from the relation:
\begin{equation}
s=
\begin{cases}
\E(e^*) - e^* & \text{if and only if } \E(e^*) \geq e^* \\
0 & \text{otherwise}
\end{cases}
\end{equation}
in such a formula $e^*$ is the number of cross-class edges (i.e. $m_{10}$) and $\E(e^*)$ is the first moment of $e^*$. Therefore, $S$ is expressed as $S = s/\E(e^*) \in [0,1]$. Any value of $S$ may be interpreted simply as the ratio of the number of missing cross-class links to the expected number of such links. By using the notation that we adopted throughout the paper, we can express the number of cross-class links as have $e^* = m_{10}$. Thus:
\begin{equation}
S = \frac{\E(m_{10})-m_{10}}{\E(m_{10})} \, .
\end{equation}
A value of $S=1$ indicates that there are no cross-class links and that segregation is complete. Whenever $m_{10}^l \neq 0$ we can guarantee the absence of complete segregation for any realisation of the considered $D_G$.

\subsection{Dataset Description}
\label{DSDescr}

\subsubsection{Facebook100}

The Facebook100 dataset \cite{traud2012social} contains an anonymized snapshot of the friendship connections among 1208316 users affiliated with the first 100 colleges admitted to Facebook. The dataset contains a total of 93969074 friendship links between users of the same college. Each node has a set of discrete-valued social attributes: status $\{undergraduate, graduate student, summer student, faculty, staff, alumni\}$, dorm, major, gender $\{male, female\}$ and graduation year.

\subsubsection{Wolf Dominance}
The network represents a set of dominance relationships among a captive family of wolves \cite{van1987dominance}. Common signs of dominance among wolves are two low postures (namely low and low-on-back) and two behaviors (namely body tail wag and lick mouth \cite{schenkel1967submission}). In such a network a node corresponds to a wolf and a link exists if a wolf exhibited a low posture to another one. 
The network with $n=16$ nodes and $m=148$ links is provided with metadata such as age and gender.
 
\subsection{Bounds for scale-free networks}
\label{BSF}

Given a certain degree sequence $D_G$ with $n$ elements that follows a power law distribution with exponent $\gamma$, the fraction of nodes holding a certain degree value $d$ is $p(d) = a d^{-\gamma}$ where $a$ is chosen so that the sum over $p(d)$ equals 1.

In order to obtain the number of nodes $n$ from the sum of $p(d)$ we simply multiply the parameter $a$ by $n$ thus:
\begin{equation}
\sum_{d_{\min}}^{d_{\max}} a d^{-\gamma} n = an \sum_{d_{\min}}^{d_{\max}} d^{-\gamma} = b \sum_{d_{\min}}^{d_{\max}} d^{-\gamma} = n \,, 
\end{equation}
where $d_{\min}$ and $d_{\max}$ are respectively the minimum and maximum degree values in $D_G$. Such a summation can be approximated by:
\begin{equation}
b \int_{d_{\min}}^{d_{\max}} x^{-\gamma} dx \,.
\end{equation}
In this continuous model of a power law distribution, where the limit of large networks is usually considered, it is a reasonable approximation to assume that the degree values extend to infinity:
 \begin{equation}
 b \int_{d_{\min}}^{+\infty} x^{-\gamma} dx \approx n \,,
 \end{equation} 
 in other words
  \begin{equation}
 b \frac{d_{\min}^{1-\gamma}}{\gamma-1}  \approx n \,.
 \end{equation} 
The maximum degree is determined by 
 \begin{equation}
b \int_{d_{\max}}^{+\infty} x^{-\gamma} dx \approx 1 \,.
  \end{equation}
since only one node has degree $\geq d_{\max}$, in other words
 \begin{equation}
 b \frac{d_{\max}^{1-\gamma}}{\gamma-1} \approx 1 \,.
 \end{equation} 
By comparison we find the classic relation:
\begin{equation}
d_{\max} \approx d_{\min} n^{1/(\gamma-1)} \,.
\end{equation}
Given a certain value of $n_1$, we can write an approximation of the tail (head) of the degree sequence $D_G^T(n_1)$ ($D_G^H(n_1)$), using arguments similar to those presented above. In order to write such an approximation we need to identify the degree value $d$ that, fixed $n_1$, guarantees us to consider $n_1$ elements of the degree sequence. Accordingly, the degree value $d$ will be the variable of our equation.

The degree value $d$ up to which we need to cut the degree sequence in order to obtain an approximation of $D_G^T(n_1)$ (made up of $n_1$ not necessarily different degree values) can be obtained via the following equation:
\begin{equation}
b \int_{d_{\min}}^d x^{-\gamma} dx \approx n_1
\end{equation}
and thus
\begin{equation}
b \int_{d}^{+\infty} x^{-\gamma} dx \approx n-n_1=n_0,
\end{equation}
in other words:
 \begin{equation}
b \frac{d^{1-\gamma}}{ \gamma-1} \approx n_0 \,.
	\end{equation}
Comparing with the above we find:
\begin{equation}
d\approx d_{\min}  (n/n_0)^{1/(\gamma-1)}  \,.
\label{eq_d}
\end{equation}
The total number of edges is $m= \sum_i d_i/2$ which can be approximated by:
\begin{equation}
m \approx \frac{b}{2} \int_{d_{\min}}^{+\infty} x x^{-\gamma} dx=
\frac{b}{2} \frac{d_{\min}^{2-\gamma}}{2-\gamma}
=\frac{n}{2} d_{\min} \frac{\gamma-1}{\gamma-2} \,.
\end{equation}
This allows to find the sum of all $n_1$ bottom degrees (i.e. of the elements in $D_G^T(n_1)$):
\begin{equation}
 \sum_{i \in D_G^T(n_1)} d_i = b \int_{d_{\min}}^d x x^{-\gamma} dx=2m - b \int_{d}^{+\infty} x x^{-\gamma} dx
\end{equation}
which is
\begin{equation*}
2m - n_0 d  \frac{\gamma-1}{\gamma-2}
\end{equation*}
that, using Eq.~\eqref{eq_d}, can be turned into
\begin{equation*}
n  d_{\min} \frac{\gamma-1}{\gamma-2} -  n_0 d_{\min}  (n/n_0)^{1/(\gamma-1)}  \frac{\gamma-1}{\gamma-2}
\end{equation*}
which can be simplified to
\begin{equation*}
n  d_{\min} \frac{\gamma-1}{\gamma-2} ( 1 -   (n_0/n)^{(\gamma-2)/(\gamma-1)}  )  
\end{equation*}
and finally to:
\begin{equation*}
\sum_{i \in D_G^T(n_1)} d_i=2m (1-  (n_0/n)^{(\gamma-2)/(\gamma-1)} ) \,.
\end{equation*}
It follows that, the sum of top degrees (i.e. of the elements in $D_G^H(n_0)$) is:  
\begin{equation*}
D_G^H(n_0) = 2m  (n_0/n)^{(\gamma-2)/(\gamma-1)}
\end{equation*}
For example, for $\gamma=3$ and $n_0=n_1=n/2$,
\begin{equation*}
\sum_{i \in D_G^T(n_1)} d_i \approx 2m (1-\sqrt{2}/2) \,.
\end{equation*}

Finally, $D^T_G(n_0)$ and $D_G^H(n_1)$ can be obtained analogously by means of the equality $n_0 = n - n_1$.
In summary:

\begin{itemize}
\item $\sum_{i \in D_G^T(n_1)} d_i =  2m (1-  (n_0/n)^{(\gamma-2)/(\gamma-1)}  )$

\item $\sum_{i \in D_G^H(n_0)} d_i=  2m  (n_0/n)^{(\gamma-2)/(\gamma-1)}  $
\item $\sum_{i \in D_G^T(n_0)} d_i=  2m (1-  (n_1/n)^{(\gamma-2)/(\gamma-1)}  )$
\item $\sum_{i \in D_G^H(n_1)} d_i=  2m  (n_1/n)^{(\gamma-2)/(\gamma-1)}  $
\end{itemize}
Since we computed the degree sums of different portions of the degree sequence analytically, the bounds to the edge counts can be simplified to:

\begin{equation}
m_{11}^u = \min\Bigg( \binom{n_1}{2}, m  (n_1/n)^{(\gamma-2)/(\gamma-1)} \Bigg)     \enspace ,
\end{equation}

\begin{equation}
m_{10}^u = \min\Bigg(n_1 n_0,  m  (n_0/n)^{(\gamma-2)/(\gamma-1)}  \Bigg) \, .
\end{equation}
(assuming $n_0 \leq n_1$)

\begin{equation}
m_{11}^l = \max \Bigg(0,  m  \bigg(   1-2(n_0/n)^{(\gamma-2)/(\gamma-1)}    \bigg)     \Bigg)
\end{equation}

\begin{equation}
m_{10}^l = \max \bigg(1; 2m \bigg(   1-2(n_0/n)^{(\gamma-2)/(\gamma-1)}    \bigg) - n_1(n_1 -1)\bigg)
\end{equation}
(assuming non trivial case $0 < n_1  < n$).

\bigskip

If we assume that $n_0/n$ is a constant fraction $\alpha_0$ (not $0$ and not $1$) while $n \to \infty$ and $d_{\min}$ and $\gamma$ remain constant, we can further simplify:

\begin{equation}
m_{11}^u =  m  \alpha_1^{(\gamma-2)/(\gamma-1)}      \enspace ,
\end{equation}

\begin{equation}
m_{10}^u =   m  \alpha_0^{(\gamma-2)/(\gamma-1)}   \, .
\end{equation}

(assuming $\alpha_0 \leq \alpha_1$)

\begin{equation}
	m_{11}^l =   m \max \Bigg( 0,      1-2\alpha_0^{(\gamma-2)/(\gamma-1)}      \Bigg)     
\end{equation}
	
\begin{equation}
	m_{10}^l = 1  \, .
\end{equation}

\begin{figure}[h]
\centering
\includegraphics[scale=0.5]{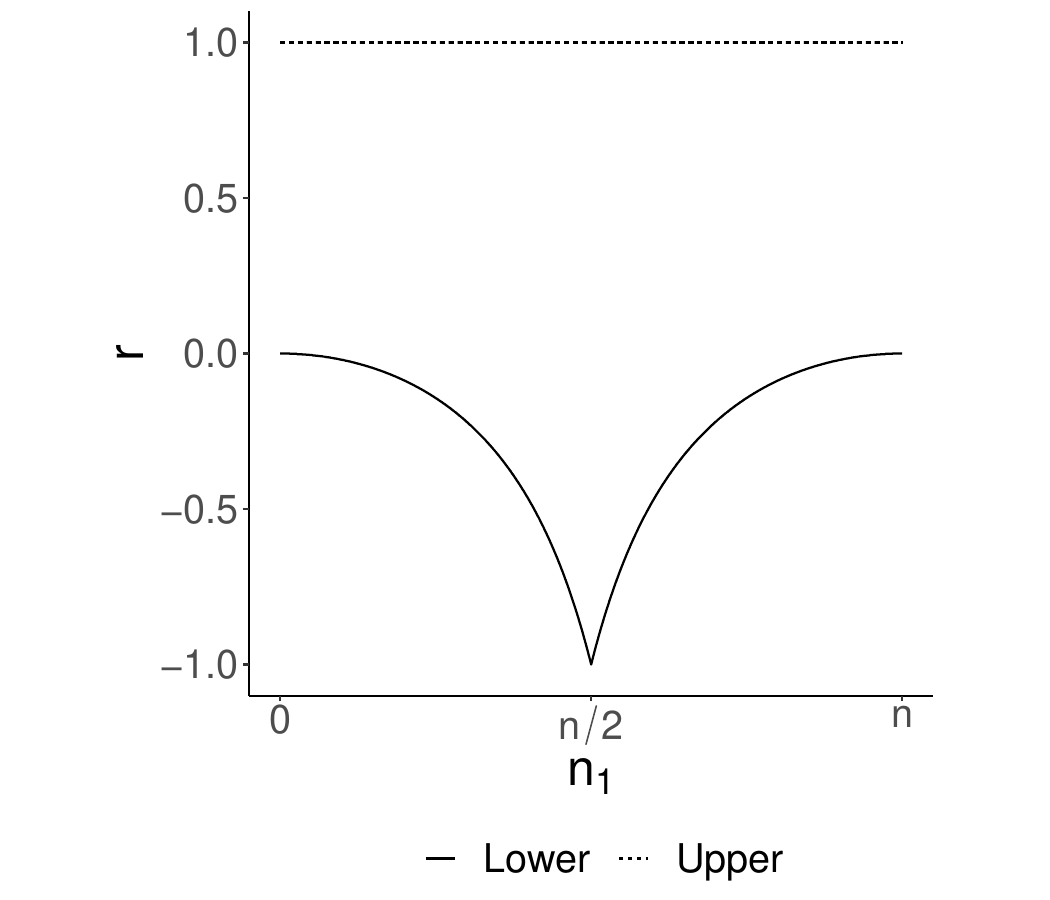}
\caption{Upper and lower bounds to binary assortativity for a scale-free network with $\gamma=3$}
\label{fig_bounds_sf3}
\end{figure}

Using the bounds to the edge counts we can compute upper and lower bounds to binary assortativity as in Eq.~\eqref{maxr} and~\eqref{min_assort}. As an example we consider a scale-free degree distribution with $\gamma = 3$ for which we obtain the bounds reported in Figure~\ref{fig_bounds_sf3}. While the upper bound in uninformative, the lower bound becomes tighter in presence of a small minority ($n_0 << n_1$ and $n_1 << n_0$) allowing room only for slightly disassortative configurations.

\end{document}